\documentclass[a4paper,11pt]{article}
\pdfoutput=1 
\usepackage{jcappub} 
\usepackage[T1]{fontenc}
\usepackage{bm,amssymb,amsfonts,slashed,amsthm,amsmath,graphicx}

\title{Efficient numerical solution to vacuum decay with many fields}

\author[1]{Ali Masoumi,}
\author[1]{Ken D. Olum,}
\author[1,2]{and Benjamin Shlaer}

\affiliation[1]{Institute of Cosmology, Department of Physics and Astronomy, 
Tufts University, Medford, MA  02155, USA}
\affiliation[2]{Department of Physics, University of Auckland, Private Bag 92019, Auckland, New Zealand}

\emailAdd{ali@cosmos.phy.tufts.edu}
\emailAdd{kdo@cosmos.phy.tufts.edu}
\emailAdd{shlaer@cosmos.phy.tufts.edu}

\def\rmax{{r_{\text{max}}}}
\def\Bmax{{B_{\text{max}}}}
\def\phitv{{\varphi_{\text{tv}}}}
\def\phifv{{\varphi_{\text{fv}}}}
\def\phimax{{\varphi_{\text{max}}}}
\def\bphi{\bm{\phi}}
\def\varbphi{\bm{\varphi}}
\def\bchi{\bm{\chi}}
\def\bphifv{{\varbphi_{\text{fv}}}}
\def\bphitv{{\varbphi_{\text{tv}}}}
\def\bphipath{{\bphi_{\text{path}}}}
\def\nn{\nonumber \\}
\newcommand{\bel}[1] {\begin{equation}\label{#1}}
\newcommand{\beal}[1] {\begin{eqnarray}\label{#1}}
\newcommand{\be}{\begin{equation}}
\newcommand{\ee}{\end{equation}}
\newcommand{\bea}{\begin{eqnarray}} 
\newcommand{\eea}{\end{eqnarray}}
\def\bx{\mathbf{x}}
\def\bd{\bm{\delta}}
\def\diag{\mathop{\text{diag}}}
\def\fb{\mathbf{f}}             
\def\bg{\mathbf{g}}
\def\bF{\mathbf{F}}
\def\bA{\mathbf{A}}

\abstract{Finding numerical solutions describing bubble nucleation is notoriously
difficult in more than one field space dimension.  Traditional shooting
methods fail because of the extreme non-linearity of field evolution
over a macroscopic distance as a function of initial conditions.
Minimization methods tend to become either slow or imprecise for larger
numbers of fields due to their dependence on the high dimensionality of
discretized function spaces.  We present a new method for finding
solutions which is both very efficient and able to cope with the
non-linearities.  Our method directly integrates the equations of motion
except at a small number of junction points, so we do not need to
introduce a discrete domain for our functions.  The method, based on
multiple shooting, typically finds solutions involving three fields in
around a minute, and can find solutions for eight fields in about an
hour.  We include a numerical package for Mathematica which implements
the method described here.}

\begin{document}
\maketitle
\flushbottom
\section{Introduction} 
\label{sec:Intro}

The thermal or quantum mechanical nucleation of bubbles initiates a first-order phase transition.  This phenomenon has application in diverse areas of physics including superfluid Helium-3 \cite{Langer:1969bc, Leggett:1975te}, Higgs-induced vacuum decay, \cite{Cabibbo:1979ay,Hung:1979dn,EliasMiro:2011aa,Dasgupta:1996qu,Sarid:1998sn}, baryogenesis and gravitational waves from a first-order electroweak phase transition \cite{Steinhardt:1981ct, Witten:1984rs, Hogan:1986qda, Kosowsky:1992rz, Cline:1999wi, Randall:2006py, Kuzmin:1985mm}, and quantum gravitational instabilities \cite{Witten:1981gj, Gross:1982cv, Randall:2006py}.   The idea that the observable universe represents but one vacuum \cite{Guth:2007ng, Vilenkin:2006xv} in a diverse landscape of string vacua \cite{Bousso:2000xa, Susskind:2003kw} motivates a study of false vacuum decay in high-dimensional field spaces \cite{Aazami:2005jf, Sarangi:2007jb, Tye:2007ja, Huang:2008jr, Greene:2013ida, Aravind:2014pva, Dine:2015ioa, Masoumi:2016eqo}.

Quantum field theories with an effective potential $U(\phi)$ have
perturbatively stable vacua corresponding to the local minima of
$U(\phi)$.  Non-perturbative effects can destabilize these local
minima via the spontaneous nucleation of bubbles containing a lower
energy perturbative vacuum \cite{Kobzarev:1974cp}.  Instanton methods
\cite{Coleman:1977py, Linde:1981zj} provide a simple and elegant
treatment of this process.  The primary calculation is a Euclidean
$O(4)$-symmetric ``bubble'' field profile (called the bounce) which
smoothly interpolates between a true-vacuum configuration at its
center and the false vacuum whose decay is being described.  Once this
is calculated, the decay rate follows, as well as the field
configuration for the formation and evolution of a single bubble.

This paper is structured as follows.  We briefly review the bounce
formalism in section~\ref{sec:Formalism}.  In
section~\ref{sec:techniques} we describe simple algorithms for finding
the bounce solution and the problems they face.  In
section~\ref{sec:Method} we describe our more robust method for
finding the bounce.  In section~\ref{sec:Examples} we give some simple
examples, and in section~\ref{sec:Quality} we describe the abilities
of our code.  In section~\ref{sec:Comparison} we compare our method
with other methods, and we conclude in section~\ref{sec:Conclusion}.

\section{$O(4)$ bounce formalism}
\label{sec:Formalism}

\begin{figure}
   \centering
   \includegraphics[width=2.8in]{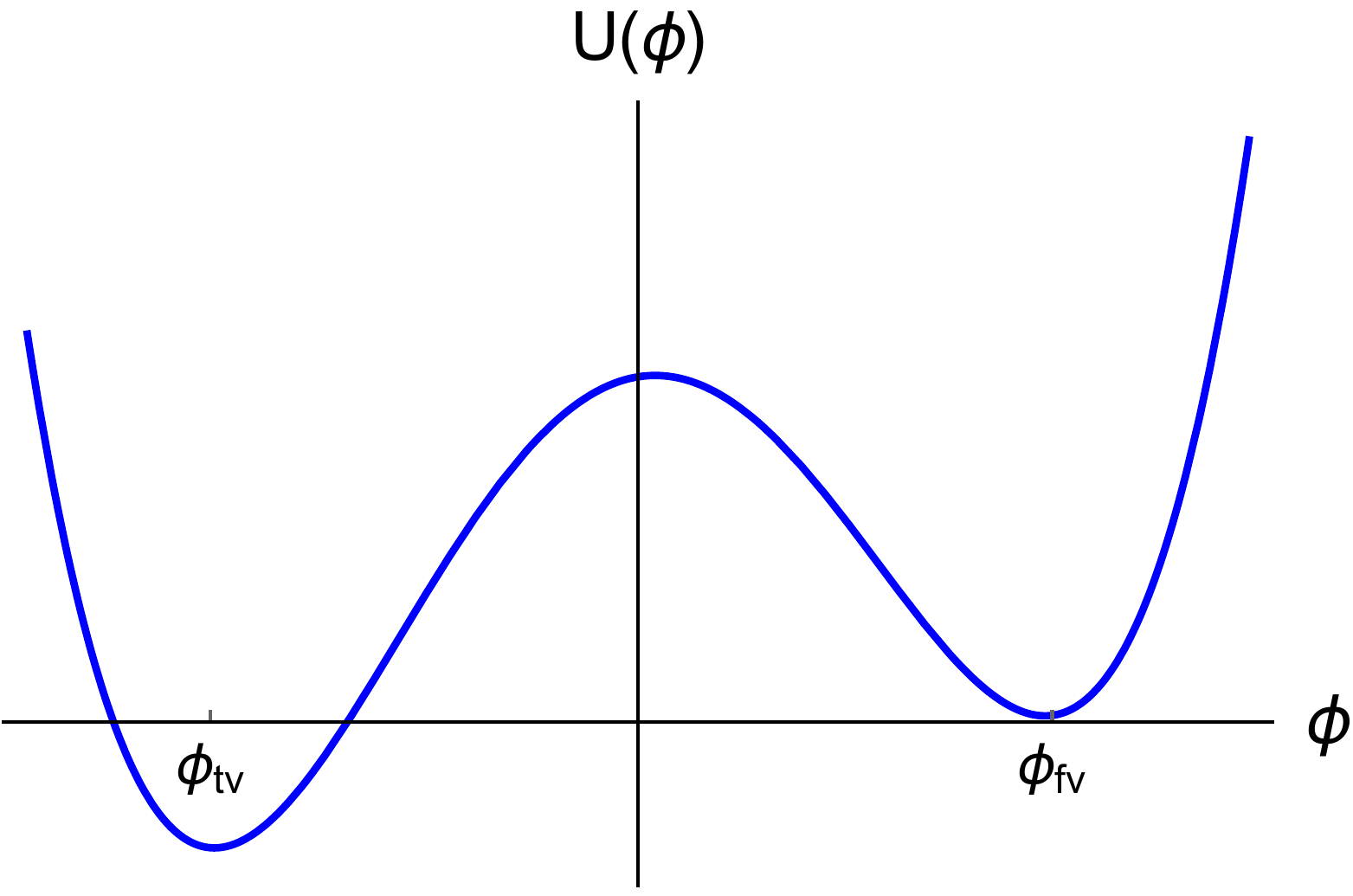} 
   \includegraphics[width=2.8in]{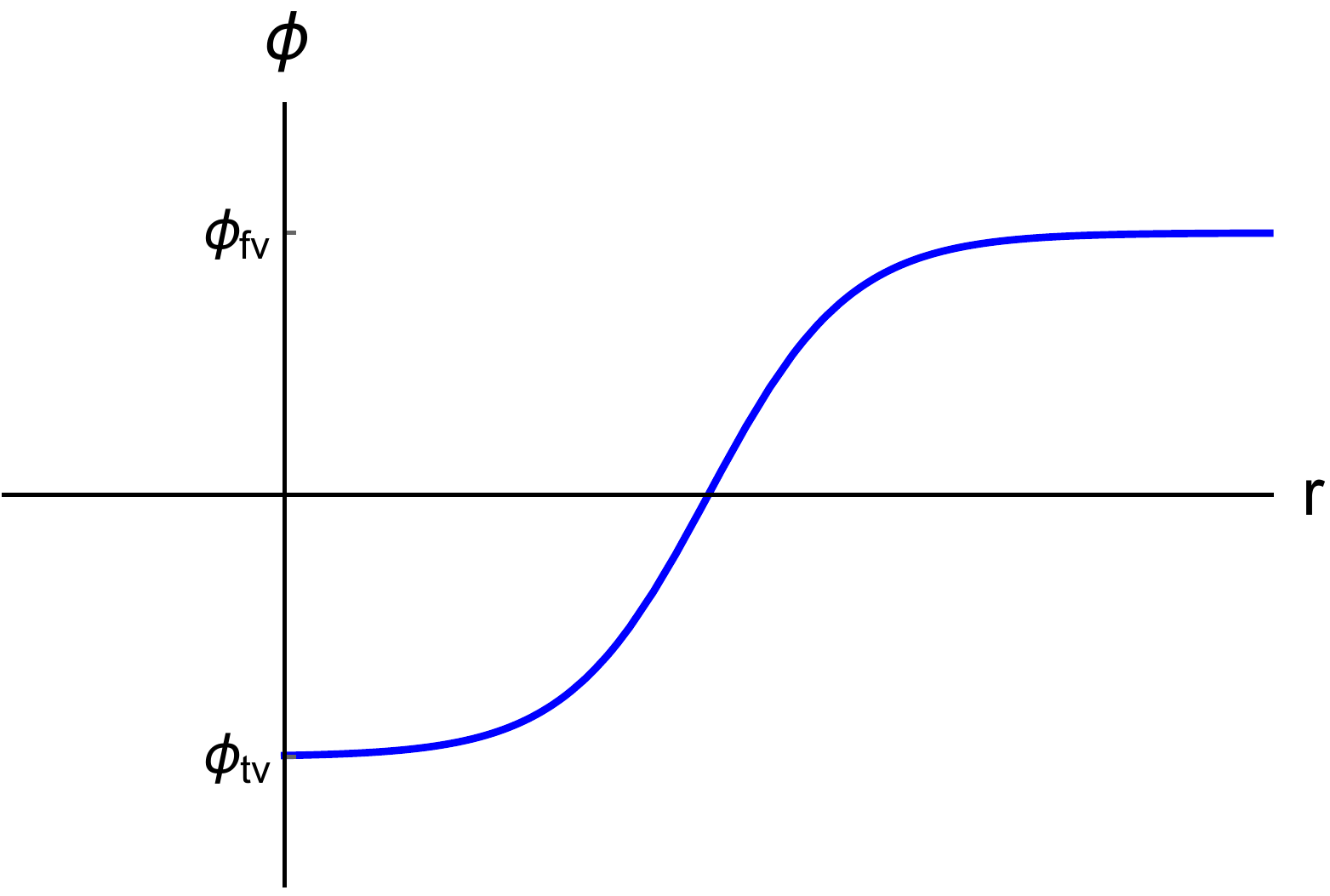} 
   \caption{Left panel shows  a potential with two vacua. The left vacuum $\phitv$ is stable and denoted as the ``true vacuum.'' The right vacuum $\phifv$, which is meta-stable, is called the ``false vacuum.'' Right panel shows the field profile that carries the tunneling. }
   \label{fig:TrueFalse}
\end{figure}

In this section we follow Coleman \cite{Coleman:1977py}. We start from
the Lagrangian for a scalar field in flat space,
\be
	{\mathcal L} = \frac{1}{2} \partial_\mu \phi \partial^\mu \phi - U(\phi)~,
\ee
where $U(\phi)$ is a potential with two minima.  An example is shown
in Figure~\ref{fig:TrueFalse}.
The tunneling is described by a solution to the Euclidean ($x_0 \to -i\tau$) equations of motion,
\be \label{eq:1DEOM}
	\nabla^2 \phi = \frac{\partial U}{\partial \phi}~.
\ee
Assuming that the field has an $O(4)$ symmetry, i.e., $\phi(x) =
\phi(r)$ where $r=\sqrt{\tau^2 + \sum x_i^2}$, \eqref{eq:1DEOM}
simplifies to
\be\label{eq:1DEOM2}
	\frac{d^2 \phi}{d r^2} + \frac{D-1}{r} \frac{d \phi}{dr} = \frac{\partial U}{\partial \phi}~,
\ee
where $D$ is the number of spacetime dimensions.  This is the equation
of motion for a particle that moves in the ``upside-down'' potential
$-U(\phi)$ under the influence of the velocity-dependent ``friction'' term
$(D-1)/{r}$. 
The boundary conditions are 
\be
\phi'(0)=0, \qquad \lim_{r\rightarrow\infty}\phi(r)=\phifv,
\ee
where the prime denotes
differentiation with respect to $r$.

 It is the friction that allows the field to roll to a stop atop the false-vacuum hill, which is lower than
the true-vacuum initial position.  Since friction weakens inversely with the time parameter $r$, beginning too close
to the peak of the true-vacuum hill will result in a large initial delay, and so too little friction when velocities are large; the field will overshoot the false vacuum.

The simplest non-constant solution with the above properties is known as the bounce.
A typical potential and the corresponding bounce field profile are
shown in Figure~\ref{fig:TrueFalse}. The tunneling rate is given by $\Gamma = A e^{-B}$,
where $B$ is the Euclidean action of the corresponding bounce minus
the action of the false vacuum configuration.

Generalization of this problem to more than one field is
straightforward. We denote the fields by $\bphi = (\phi_1, \phi_2, \ldots,
\phi_N)$ and the potential by $U(\bphi)$. The bounce is then a
solution to the system of coupled ordinary differential equations
\be\label{eq:MultiFieldDEOM}
	\frac{d^2 \phi_i}{d r^2} + \frac{D-1}{r} \frac{d \phi_i}{dr} = \frac{\partial U}{\partial \phi_i}~,
\ee
with boundary conditions $d\bphi/dr |_ {r=0}=0$ and $\lim_{r
  \rightarrow \infty} \bphi(r) =\bphifv$.  This is a
system of $N$ second-order equations with $N$ boundary conditions at $r=0$ and
another $N$ at $r=\infty$. The field must begin at $r=0$ from some specific
point at the center of the bubble and asymptotically approach the false
vacuum infinitely far away.  Finding the solution to the bounce
equations is tantamount to determining the values of the fields at $r=0$.

\begin{figure}
\centering
\includegraphics[width=3in]{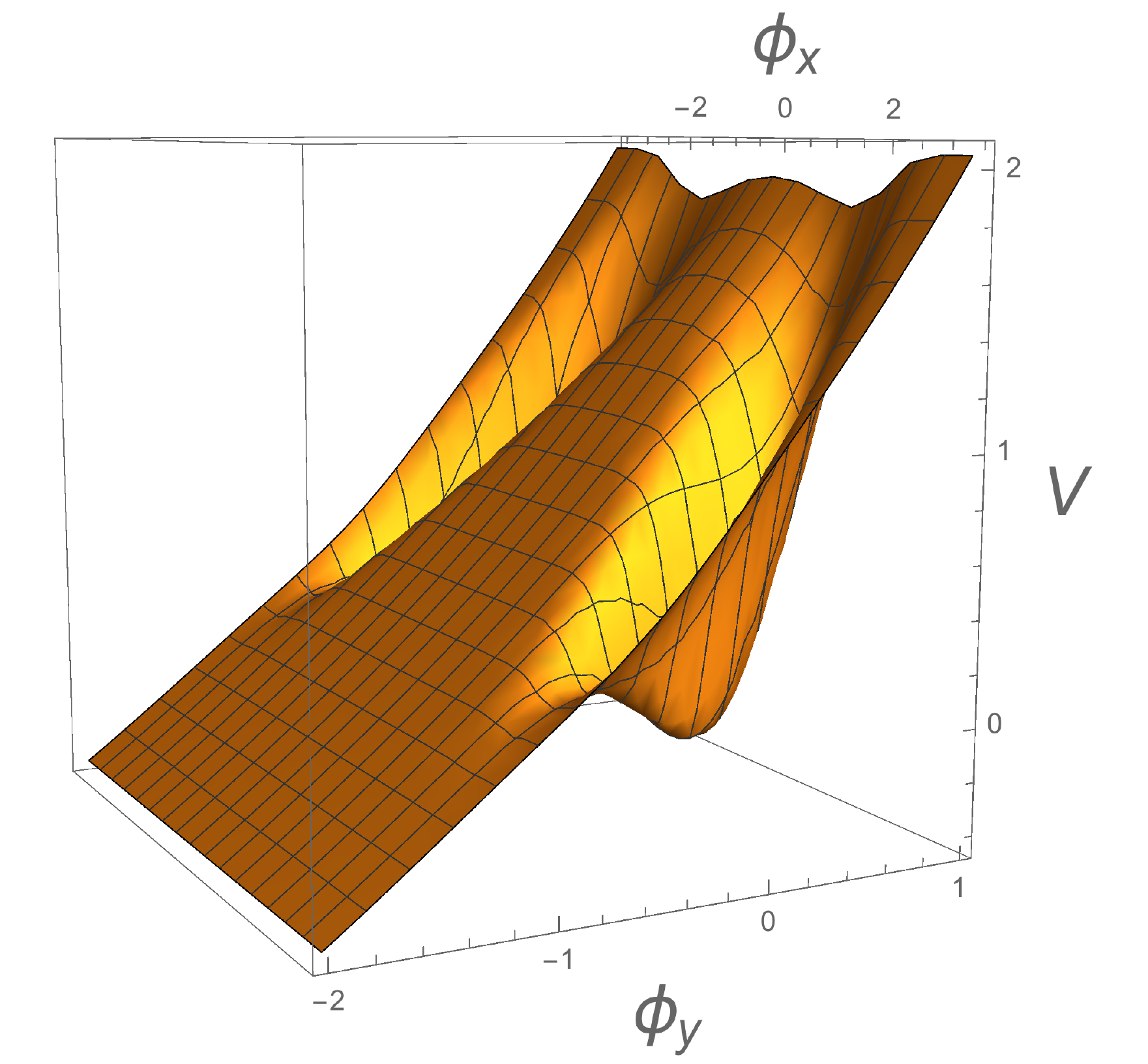} 
\caption{In this two-dimensional potential there is no instanton
  connecting the two minima on the right.  The problem is that
  tunneling between the two minima is dominated by tunneling to lower
  values on the left.}
\label{fig:nosolution}
\end{figure}

For a single field, Coleman \cite{Coleman:1977py} used an
overshoot-undershoot argument to prove that there is always a bounce
that connects a higher meta-stable vacuum to an adjacent lower energy
vacuum.  However, with more than one field there are cases where no
solution exists, as shown, for example, in
figure~\ref{fig:nosolution}.

\begin{figure}
   \centering
   \includegraphics[width=2.5in]{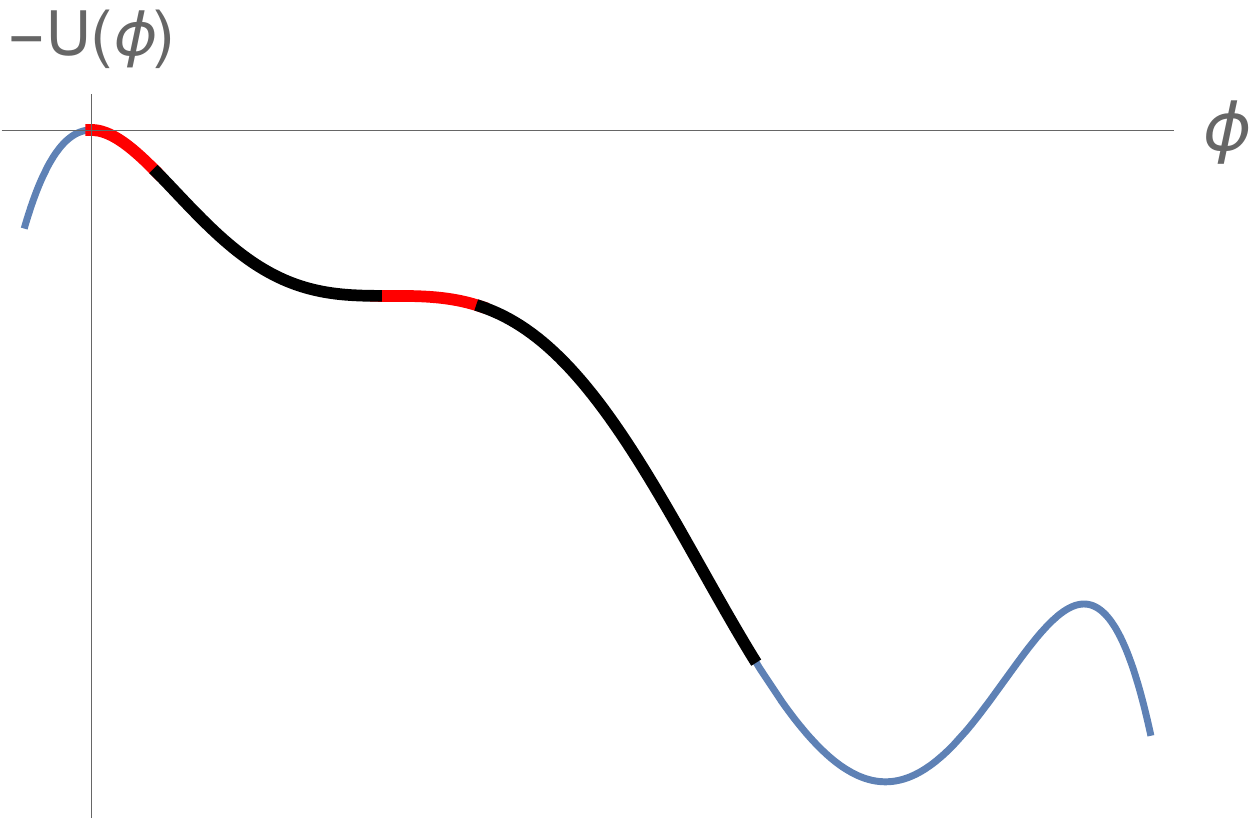} 
   \caption{A one-dimensional potential, shown inverted, that has
     more than one solution.}
   \label{fig:OddPotential}
\end{figure}

There are also cases that have more than one solution.  With more
than one field, there may be more than one ``mountain pass''
connecting the two vacua, as in figure~\ref{fig:2dExample} below.
Each pass gives rise to its own instanton.  Even with only one field,
there are cases with multiple solutions. One example is shown in
figure~\ref{fig:OddPotential},
where we drew the upside down potential.  Regions shown in red
overshoot the true vacuum, while those shown in black undershoot.
Thus the right edge of the upper red region and the two edges of the
lower red region are each a solution.

For a thermally induced first-order phase transition the formalism is 
quite similar \cite{Langer:1969bc, Linde:1981zj}.

\section{Solution techniques}\label{sec:techniques}

The most straightforward method to find the bounce numerically would
be shooting from the initial condition $\bphi(0) = \varbphi_0$.  Starting
from some guess for $\varbphi_0$, we could integrate the equations of
motion to large $r$. The correct value of $\varbphi_0$ yields the desired
asymptotic boundary condition $\bphi \to \bphifv$.  We could
iteratively improve $\varbphi_0$ based on how close $\bphi$ comes to $\bphifv$.

This technique requires integrating the differential equation
\eqref{eq:MultiFieldDEOM} from $r=0$ to large $r$.  For a single field, so
long as we are descending the inverted potential from the true vacuum,
this equation is stable.  But as we ascend the inverted potential
toward the false vacuum, it becomes unstable.  The desired solution
asymptotically approaches $\phifv$, but there is another solution which
more and more rapidly falls down the hill.  This growing mode causes
numerical instability that makes it impossible to find the solution
with the correct $\phi \to \phifv$ boundary condition.

This problem can be avoided by a ``double shooting'' scheme \cite{Bayliss}.
We pick $\phi_0$ as above and pick also some large but finite value
$\rmax$ and a field value $\phi(\rmax) = \phimax$ there.  Then we
integrate \eqref{eq:1DEOM2} to the right from $r = 0$ and to the left
from $\rmax$, and demand that the two solutions match at some
intermediate location.  We then attempt to iteratively improve $\phi_0$
and $\phimax$.

The double shooting technique works reasonably well for one field, but
for additional fields we face additional problems.  One problem is
that the solution may travel along a ``ridge'' in the inverted
potential.  In this case, there is an unstable mode that falls off the
ridge instead of traveling along it.  Reversing the direction of
integration does not help this problem.

This problem exists even when the field is descending the inverted
potential.  Unless it is following the steepest descent, modes that
descend in steeper directions will always be more unstable than the
desired mode.  This occurs even near the true and near the false
vacuum.

A related problem is that in many cases, $\varbphi_0$ can never be
specified with sufficient accuracy.\footnote{The one-field
  version of this problem is discussed by Wainwright
  \cite{Wainwright:2011kj}, who solves it by using the logarithm of
  the distance from the true vacuum as a parameter.}  For example,
consider a potential which near the true vacuum has the form
\be
U(\phi_1,\phi_2) = 10(\phi_1^2+\phi_2^2) - 16\phi_1\phi_2.
\ee
Neglecting friction, the solution is
\bea
\phi_1 &=& A e^{2r} + B e^{6r},\\
\phi_2 &=& A e^{2r} - B e^{6r}.
\eea
Suppose we want to reach $\bphi=(2,0)$ at $r=10$.  Then we
need $A = e^{-20}, B = e^{-60}$, and thus
$\varbphi_0=(e^{-20}+e^{-60},e^{-20}-e^{-60})$. But in double-precision
floating point, $e^{-20}\pm e^{-60}$ is indistinguishable from
$e^{-20}$, so we end up with $\varbphi_0=(e^{-20},e^{-20})$, and thus
$\bphi(10) = (1,1)$ instead of $(2,0)$.  Including the friction term
would lead to Bessel functions instead of exponentials, but the effect
is the same.

\section{Description of the method}
\label{sec:Method}

As mentioned in section~\ref{sec:Formalism}, the simple shooting
methods for solving the bounce equations face several difficulties,
which we address here. The most significant problem is solving a set
of differential equations that are unstable. We overcome this problem
using a multi-interval shooting method described in
subsection~\ref{subsec:Multi-Shooting}. At the center of the bubble
($r=0$) there is a divergence in
\eqref{eq:MultiFieldDEOM}. Additionally, the other boundary condition
is set at $r=\infty$, which cannot be represented as a floating-point
number.  We solve these problems by using an approximate analytic
solution in subsection~\ref{subsec:Analytic}. In
subsection~\ref{sec:AddingNewVar} we add an additional variable to the
set for which we are solving, in order to make it easier for the
system to change the solution by translation in $r$, in
subsection~\ref{sec:Rescaling} we rescale the potential so that all
function values used in the solution have similar ranges, and in
subsection~\ref{sec:Readjustingvalues} we make changes to the division
points between the shooting regions.

To find the bounce solution using shooting methods we need to come up
with an initial guess. The procedure for finding a good initial guess
is described in subsection~\ref{subsec:InitialProfile}.  After we find
the solution, we need to calculate the action, which is discussed in
subsection~\ref{sec:Derrick}.

Solving non-linear equations using steepest decent guarantees that the
steps taken toward the solution are small, but this method converges
slowly. On the other hand Newton's method converges rapidly, but can
take dangerously large steps. We use Powell's hybrid method, which
converges quickly near the solution, yet takes reasonable steps far
away from it. This is explained in appendix~\ref{sec:Powell}. To use
the this method we need to calculate the Jacobian of the field values
with respect to the initial conditions.  We use an accurate method for
calculating the Jacobian which is described in
appendix~\ref{sec:Jacobian}.  Appendix~\ref{sec:UsingTheCode} explains
how to download and run the code.

\subsection{Multiple shooting method}
\label{subsec:Multi-Shooting}
The differential equations \eqref{eq:MultiFieldDEOM} are unstable in the
sense that there are growing and decaying modes, and the growing
modes may have different rates of growth.  Unless we are interested in
the fastest mode, a small admixture of this mode from numerical error
will eventually overpower the solution of interest.  To avoid this
problem, we integrate the differential equations only over a range of
$r$ small enough that the growth is tolerable, and put together
several such regions to cover the needed range of $r$.  This method is
known as multiple shooting
\cite{Morrison:1962:MSM:355580.369128}.

To understand the degree of instability, let us expand $U(\bphi)$
around some specific point $\varbphi_1$ to quadratic order,
\be\label{eq:Taylor1}
U(\bphi) = U(\varbphi_1) + A_i(\bphi - \varbphi_1)_i+ (B_{ij}/2)(\bphi -\varbphi_1)_i(\bphi -\varbphi_1)_j + \mathcal{O}(\bphi-\varbphi_1)^3~,
\ee
where 
\be \label{eq:ABDef}
A_i= \left.\frac{\partial U}{ \partial \phi_i}\right|_{\varbphi_1}~, \qquad B_{ij}=  \left.\frac{\partial^2 U }{\partial \phi_i \partial \phi_j}\right|_{\varbphi_1}~.
\ee
If we use \eqref{eq:ABDef} and ignore the frictional term in
\eqref{eq:MultiFieldDEOM}, we have
\be
	\frac{d^2 \phi_i}{d r^2}= A_i + B_{ij} (\bphi-\varbphi_1)_j~.
\ee
We can diagonalize the symmetric matrix $B_{ij}$ using an orthogonal
matrix $O$, so $B_{ij} = O_{ki} \tilde B_{kl} O_{lj} $ with $\tilde
B_{kl} = \diag(B_1,  B_2, \ldots, B_N)$.  Define
$\tilde\phi_i = O_{ij} (\bphi- \varbphi_1)_j$ and $\tilde A_i = O_{ij}
A_j$, so that
\be\label{eq:simplifiedEOM}
	\frac{d^2 \tilde\phi_i}{d r^2} = \tilde A_i + B_i  \tilde\phi_i ~.
\ee
Assuming that all $B_i$ are positive, which is the case near the true
and false vacua, but not everywhere, the solutions to
\eqref{eq:simplifiedEOM} include a mode that grows as $e^{\sqrt{B_i}}$
and one that shrinks as $e^{-\sqrt{ B_i}}$.  If $\Bmax$ is the largest
of the $B_i$, then the worst case is that the correct solution goes as
$e^{-\sqrt{\Bmax}}$, so the fastest-growing mode grows relative to the
correct solution as $e^{2\sqrt{\Bmax}}$.

\begin{figure}
    \centering
    \includegraphics[width=2.7in]{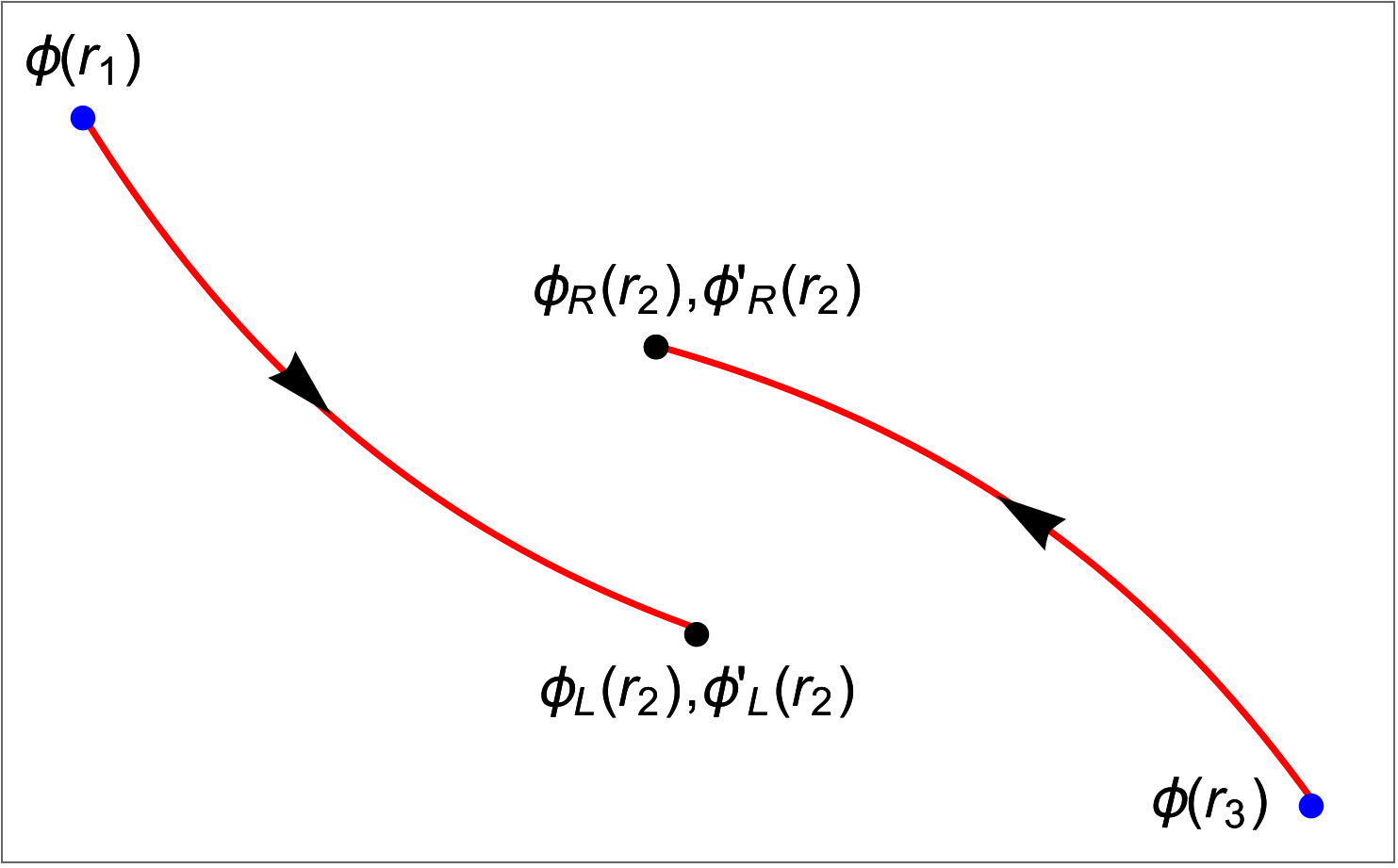} 
    \includegraphics[width=2.7in]{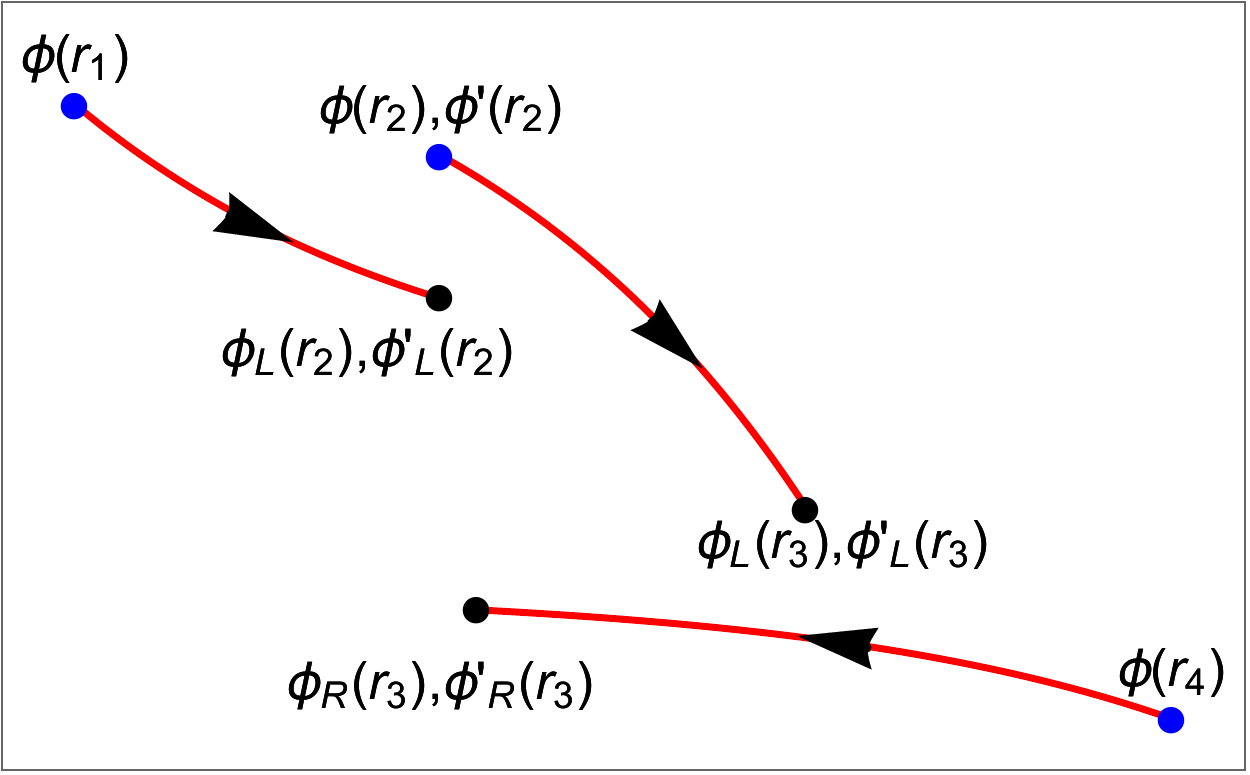}  
    \caption{The left panel shows the parameters used when there are
      only three points. We try to find $\bphi(r_1)$ and $\bphi(r_3)$ to
      make $\bphi_L(r_2)=\bphi_R(r_2)$ and
      $\bphi'_L(r_2)=\bphi'_R(r_2)$.  The right panel shows the case of
      four points. Here we try to find $\bphi(r_1)$,
      $\bphi(r_2)$, $\bphi'(r_2)$, and  $\bphi(r_4)$ to make
      $\bphi_L(r_{2})=\bphi(r_{2}), \bphi'_L(r_{2})=\bphi'(r_{2})$,
      $\bphi_L(r_3)=\bphi_R(r_3)$, and $\bphi'_L(r_3)=\bphi'_R(r_3)$.}
    \label{fig:schematics}
\end{figure}

Let the domain $r \in [0,\infty)$ be divided by $n\ge 3$ intermediate
points, $\{r_1, r_2, \ldots ,r_n\}$.  We discuss the method of
choosing these points in subsection~\ref{subsec:InitialProfile}.  At
each step, our variables are the values of the field at $r_1$ and
$r_n$ and for $n>3$ the field and its derivative at $r_2,
\ldots,r_{n-2}$.  Notice that we do not use the value of the field and
its derivative at $r_{n-1}$ as variables.  As described in the next
subsection, we do not solve the differential equation for $r<r_1$ nor
$r>r_N$.  Instead, in these regions, we approximate the potential by a
quadratic and so get an analytic solution for $\bphi(r)$, and using
this solution we determine $\bphi'(r_1)$ and $\bphi'({r_n})$. Having
the field and derivatives we can do the following numerical
integrations of \eqref{eq:MultiFieldDEOM}:
\begin{enumerate}
	\item Integrate from $r_i$ to $r_{i+1}$ for $i=1\ldots n-2$ to
          obtain $\bphi_L(r_{i+1})$ and $\bphi'_L(r_{i+1})$.
	\item Integrate from $r_n$ to $r_{n-1}$ to
          $\bphi_R(r_{n-1})$ and $\bphi'_R(r_{n-1})$.
\end{enumerate}
Here $\bphi_L$ and $\bphi_R$ denote the values of these quantities
obtained by integrating from the left and from the right, respectively.
Since we want a smooth solution to the equations of motion, we try to
match the values of $\bphi_L$ and $\bphi'_L$ with the corresponding
variables at $r_2, r_3, \ldots r_{n-2}$ and $\bphi_L$ and $\bphi'_L$
with $\bphi_R$ and $\bphi'_R$ at point $r_{n-1}$.   This is shown
for $n=3$ and $n=4$ in Figure~\ref{fig:schematics}.

The problem is now reduced to finding the values of $(2n-4)N$
variables that satisfy $(2n-4)N$ equations. We use Powell's hybrid
method, described briefly in appendix \ref{sec:Powell}, for this
purpose.

The reason for shooting toward the left in the last interval is that
in this interval (at least by $r_n$) the inverted potential is
increasing toward the false vacuum.  Shooting forward corresponds to
going uphill which is unstable in proportion to the largest eigenvalue
around the false vacuum, but shooting downhill is only sensitive to
the difference between eigenvalues, and hence we reduce the sensitivity
of the solution.  A possible improvement would be to shoot toward the
left for more intervals if the potential is also increasing there, but
we don't have any way to know in advance which these will be, and so
we did not implement this technique.

\subsection{Analytic solutions }
\label{subsec:Analytic}
As described above, we use closed-form solutions to equations with an
approximate potential for $r<r_1$ and $r>r_N$.  This technique solves
the problem where finite numerical accuracy makes it impossible to
solve the equations of motion at $\bphi = \varbphi_0$, and another problem 
where if we integrate the
differential equation for sufficiently large $r$, numerical inaccuracy
will always lead to a divergence.

\subsubsection{Analytic solution for large $r$}
\label{subsubsec:FarOut}
When the field is very close to the false vacuum, we can approximate
$U(\bphi)$ with a quadratic.  This allows an approximate analytic
solution for $\bphi$.  Of the two solutions to the second order field
equations, we clearly want the one which approaches a constant as $r
\to \infty$.  This enables us to deduce $\bphi'(r_n)$ as a function of
$\bphi(r_n)$.  We then use $\bphi(r_n)$ and $\bphi'(r_n)$ to integrate
backward to $r_{n-1}$ and apply the matching conditions there.

So consider $r_n$ large enough so that the fields $\bphi(r_n)$ are
very close to their false vacuum values. Unless the field has the
correct derivative, it will fall off the hill in some direction; for a
given $\bphi(r_n)$ there is a unique $\bphi'(r_n)$ that lands on the
hilltop.  The potential is given by the expansion \eqref{eq:Taylor1}
around $\bphifv$, with $A_i = 0$ here, and the equation of motion
becomes
\be\label{eq:MultiFieldDEOM5}
	\frac{d^2 \tilde\phi_i}{d r^2} + \frac{D-1}{r} \frac{d \tilde\phi_i}{dr} =  B_i  \tilde\phi_i ~.
\ee
We are interested in the solution which approaches $\tilde\phi_i(r)=0$
as $r \to \infty$, which is
\be
	\tilde\phi_i(r)= C_ir^{\frac{2-D}{2}} K_{\frac{D-2}{2}}\left( \sqrt{B_i} r\right)~,
\ee
where $K_{(D-2)/2}$ denotes the Bessel function of type $K$, and
the $C_i$ are constants. Returning to the original field variables,
\be\label{eq:MultiFieldDEOM4}
	\phi_i(r) =  \phifv_i+ \sum_j O_{ji} C_jr^{\frac{2-D}{2}} K_{\frac{D-2}{2}}\left( \sqrt{B_j} r\right)~.
\ee
For a given $\bphi(r_n)$ we determine the $C_i$ and from there
calculate $\bphi'(r_n)$, which we use to shoot towards $r_{n-1}$.

\subsubsection{Analytic solutions near the center of the bounce}
\label{subsubsec:NearCenter}
The situation near $r = 0$ is similar.  We do not use $\varbphi_0$
as a parameter, but instead use $\varbphi_1 = \bphi(r_1)$.  Here we
\emph{do not} assume that $\varbphi_1$ is close to the true vacuum, and
we expand not around $\bphitv$ but around $\varbphi_1$, which is accurate
as long as the $\varbphi_0$ that we infer is not too far from $\varbphi_1$.

The expansion is exactly as in \eqref{eq:Taylor1}, and the resulting
equation of motion is
\be\label{eq:MultiFieldDEOM6}
	\frac{d^2 \tilde\phi_i}{d r^2} + \frac{D-1}{r} \frac{d \tilde\phi_i}{dr} = \tilde A_i + B_i  \tilde\phi_i ~.
\ee
The initial condition $\tilde\phi'_i(0)=0$ picks out the solution with
$I$-type Bessel functions,
\be
\tilde\phi_i(r)= C_ir^{\frac{2-D}{2}} I_{\frac{D-2}{2}}\left(
\sqrt{B_i} r\right)-\frac {\tilde A_i} {B_i}~,
\ee
and the requirement that $\bphi(r_1)=\varbphi_1$, so $\tilde\phi_i(r_1)=
0$, fixes the $C_i$, giving
\be
	\tilde\phi_i(r)= \frac {\tilde A_i} {B_i} \left[ \left(\frac{r_1}{r}\right)^{\frac{D-2}{2}}\frac{I_{\frac{D-2}{2}}(r\sqrt{B_i} )}{I_{\frac{D-2}{2}}(r_1\sqrt{B_i} )}-1\right]~.
\ee
In the original field variables,
\be
\phi_j(r) = \phi_{1j} + \sum_j O_{ji}\frac {\tilde A_j} {B_j} \left[ \left(\frac{r_1}{r}\right)^{\frac{D-2}{2}}\frac{I_{\frac{D-2}{2}}(r\sqrt{B_j} )}{I_{\frac{D-2}{2}}(r_1\sqrt{B_j} )}-1\right]~.
\ee
Thus from $\varbphi_1$ we can find the derivative $\bphi'(r_1)$, and we
use these to shoot to $r_2$.

\subsection{Adding a new variable}
\label{sec:AddingNewVar}
\begin{figure}
    \centering
    \includegraphics[width=3in]{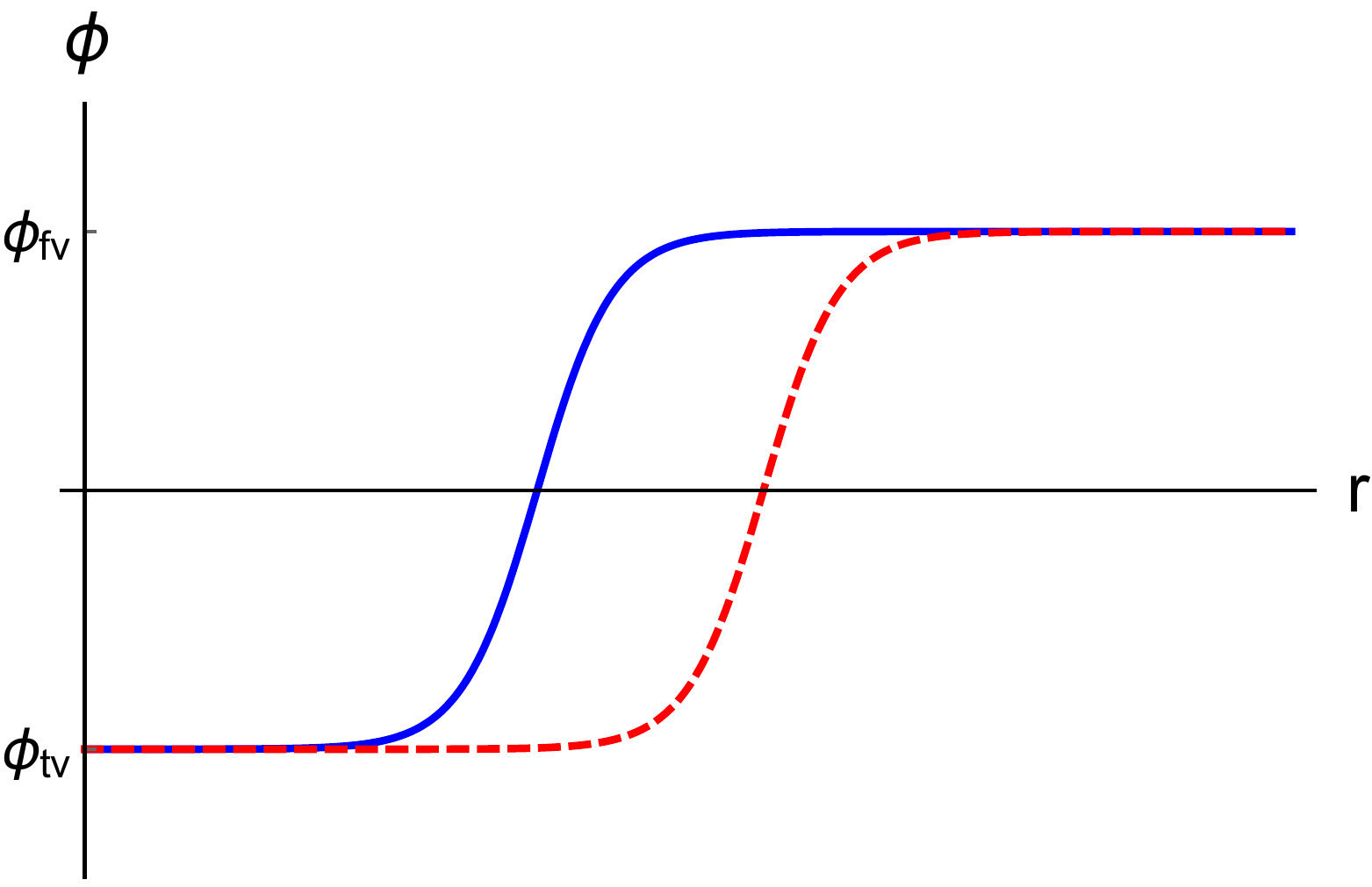} 
    \caption{Correct (solid blue) and translated (dashed red) field
      profiles.  If the code finds the translated profile, it may be
      difficult for it to find the correct profile.}
    \label{fig:shiftInv}
 \end{figure}

A generic field profile of the correct bounce solution (e.g.,
figure~\ref{fig:TrueFalse}) has plateaus at small and large $r$ and a
relatively small transition region.  However, we do not have any a
priori knowledge of the $r$ values near the transition. This leads to an important
numerical difficulty. The differential equations in
\eqref{eq:MultiFieldDEOM} exhibit an approximate $r$-translation
symmetry that is broken by the friction term. If $\bphi(r)$ is a
solution, $\bphi(r+\Delta)$ would almost satisfy the differential
equation and we will be in a situation depicted in
Figure~\ref{fig:shiftInv}.
If the code finds a translated version of the field profile, it nearly
satisfies the constraints.  While the two profiles are simply related
by translation in  $r$, in the field variables used in the solution,
$\phi(r_1), \phi(r_2), \phi'(r_1), \ldots$ the translation is very
complicated, requiring adjustment of all variables in a coordinated
way.  This makes it difficult for the code to make progress after
finding the translated profile.  A similar issue was discussed in
Ref.~\cite{Konstandin:2006nd}

To solve this problem, we add to the variables used by the equation
solver an additional variable $\Delta$, representing a translation of
the parameters $r_i$.  Instead of using the set of points $r_1, r_2,
\ldots, r_n$ shown in Figure~\ref{fig:schematics} we use the set of
$r_1+\Delta, r_2+\Delta, \ldots, r_n+\Delta$.  Thus changing $\Delta$
does not change the values of the fields and derivative variables, but
changes the location where these variables are applied, making it
simple for the solver to move in the direction of translations in $r$.

By introducing $\Delta$, we have one more unknown than our equations,
so there are infinitely many solutions, but any one is the solution to
the problem, expressed in its own way. This method proves to be very
effective and extends the range of potentials for which the code can
find solutions.  This technique is discussed further in
Ref.~\cite{Olum:2016svf}.
 
\subsection{Rescaling}
\label{sec:Rescaling}
The bounce solution has a simple transformation under rescaling of the
fields and the potential.  If the potential $U(\bphi)$ has the bounce
solution $\bphi(r)$, then the potential $V(\bchi) = \alpha
U(\bchi/\beta)$ has the bounce solution $\bchi =
\beta\bphi(\sqrt{\alpha}r/\beta)$.  The solution is also invariant
under a constant offset of the potential.

We take advantage of these transformations to modify the potential and
the field so that the distance between the true and the false vacua is
1, the value of the potential at the true vacuum is 0, and the maximum
value reached by the potential along the initial path from the true to the
false vacuum is 1.  These choices mean that the values of the field and
its derivatives that we try to match are all of the same order,
which makes it easier to find the solution.

We also change the scale of the variable $\Delta$
\cite{Olum:2016svf} by offsetting the $r_i$ by $s \Delta$ rather
than just $\Delta$.  We choose $s$ to be the average of the $r_i$
values.  This makes the magnitude of $\Delta$ similar to that of the
other parameters, even in extreme thin-wall cases with $r_i \gg 1$.

\subsection{Readjusting $r$ values}
\label{sec:Readjustingvalues}
We used approximate analytic solutions in
sections~\ref{subsubsec:FarOut} and \ref{subsubsec:NearCenter}. For
these approximations to be valid, we need $r_1$ small enough and $r_n$
large enough so that $\varbphi_1$ is close to $\varbphi_0$ and $\varbphi_n$
close to $\bphifv$.  While we construct the initial guess (see section
\ref{subsec:InitialProfile}) so that this is true, later improvement
of $\varbphi_1$ and $\varbphi_n$ may move these values outside the range
that allows for an accurate approximation.  Conversely, if $\varbphi_1$
is very close to $\varbphi_0$ or $\varbphi_n$ very close to $\bphifv$, it
means that we can use the analytic solution for larger regions.

To keep $\varbphi_1$ and $\varbphi_n$ in the desirable ranges, we use a
constant, $f_s$, giving the permissible distance between $\varbphi_0$ and
$\varbphi_1$ and another, $f_e$, giving the permissible distance between
$\varbphi_n$ and $\bphifv$, both as a fraction of the distance from
$\bphitv$ to $\bphifv$.  Both constants default to 0.01.  If the code
reaches distances that are larger, we add new $r$ points below the
current $r_1$ or above the current $r_n$ to the points $r_1, \ldots,
r_n$, and if they are much smaller (by factor 10), we remove points
from $r_1, \ldots r_n$ and use the analytic solution over a larger
range.

\subsection{Choice of the initial profile}
\label{subsec:InitialProfile}
Probably the most important step in solving a set of equations is
choosing an initial point for the solver which is close to the real
solution.  A bad choice which is out of the basin of attraction would
make it impossible to find the solution.  When we have more than one
field, there are two parts to the initial guess: what path through
field space should the initial profile follow, and how should it move
along this path as a function of $r$?

For the path, by default we simply take a straight line from the true
to the false vacuum.  But we also allow specification of a set of points
$\{\varbphi_1, \ldots, \varbphi_P\}$ through which the initial solution must go.
In this case we draw a smooth curve (a cubic spline, or quadratic if
$P=1$) through the points $\{\bphitv, \varbphi_1, \varbphi_2, \ldots, \varbphi_P,
\bphifv\}$.  We parameterize this path as  $\bphipath(\lambda)$ with $\lambda = 0\ldots1$.

For the shape of the profile as a function of $r$, we use an initial
guess based on a thin-wall analogy.  We first define a one-dimensional
potential
\be
	U_{\text{1d}}(\lambda)= U(\bphipath(\lambda))+ (3\lambda^4-4 \lambda^3)(U(\bphifv)-U(\bphitv)) ~.
\ee
This modified potential has two degenerate minima at $\lambda = 0$ and
$\lambda = 1$ and it shares the same curvature with
the original potential at the true vacuum. This potential allows for a
domain-wall solution in the form 
\bel{eq:initFieldProfile}
 r(\lambda)= r_0+ \int \frac{d\lambda}{\sqrt{2
    U_{\text{1d}}(\lambda)}} \left| \frac{d \bphipath}{d\lambda}
\right|~.
\ee
We can compute the surface tension of this solution,
\bel{eq:surfacetension}
\sigma = \int d\lambda \,\sqrt{2 U_{\text{1d}}(\lambda)}
 \left| \frac{d \bphipath}{d\lambda}
\right|~,
\ee
and determine the bubble radius in the thin-wall limit by setting the
the energy of the bubble wall according to
Eq.~(\ref{eq:surfacetension}) equal to the decrease in energy from the true
vacuum inside, giving
\be
r_{\text{thin}} = \frac{(D-1)\sigma}{U(\bphifv) - U(\bphitv)}
\ee

We chose the constant of integration $r_0$ so that the field at the
half-way path length between true and false vacua is at
$r_{\text{thin}}$. The initial profile is then defined by inverting
\eqref{eq:initFieldProfile} and using $\bphi_{\text{initial}}(r) =
\bphipath(\lambda(r))$.

Once we have the profile, we need to choose the points $r_i$.  We
choose $r_1$ to be midway from the center to the edge of the region
where the analytic approximation can be used, as defined in
section~\ref{sec:Readjustingvalues}, and similarly $r_n$ midway from
the false vacuum to the edge of the final analytic region.

To decide how many $r_i$ values to use, we attempt to make the growth
of the field during any shooting region no more than a certain limit,
by default a factor of 30.  We make the approximation of exponential
growth as in section~\ref{subsec:Multi-Shooting}.  Thus the $r$ values
should be separated by no more than $\delta\rmax$ where
$e^{\delta\rmax\sqrt{\Bmax}}$ is the allowable growth and $\Bmax$ the
largest eigenvalue of the Hessian matrix.  However, since we don't
know what values of the field we will use, we find the eigenvalues of the Hessian in the true and false vacua
and choose $\Bmax$ to be the largest of any of these.
Having found $\delta\rmax$, we divide the range between $r_1$ and $r_n$
into the smallest number of equal segments that leads to an interval no
more than $\delta\rmax$.

\subsection{Calculation of the action}
\label{sec:Derrick}
Once we have found the correct profile for the instanton, we would
like to find the tunneling action.  It is given by 
\bel{eq:action1}
	S[\bphi]= A_{D-1} \int_0^\infty dr r^{D-1}
\left[\frac12 \partial_r\phi_i\partial_r\phi_i + U(\bphi(r))-U(\bphifv)\right]=S_1+S_2~,
\ee
where $A_{D-1}$ is the area of the unit (D--1)-dimensional sphere,
and $S_1$ is the kinetic term and $S_2$ the rest.  Assuming that
$\bphi(r)$ is a solution to the equations of motion, we can transform
\eqref{eq:action1} in several ways.  First, consider a family of
scale-transformed functions $\bphi(\Lambda
r)$. We can calculate the action in \eqref{eq:action1} using
the profile $\bphi(\Lambda r)$, following \cite{Derrick:1964ww},
\bel{action2} S(\Lambda)= A_{D-1} \int_0^\infty dr r^{D-1}
\left[\frac12 \partial_r \phi_i(\Lambda r)\partial_r
 \phi_i(\Lambda r) + U(\bphi(\Lambda r))-U(\bphifv)\right]
=\Lambda^{2-D} S_1+ \Lambda^{-D}S_2~,
\ee
This action must have a stationary point at $\Lambda=1$, which
implies
\be\label{eq:Derrick}
D S_2= S_1(2-D)~.
\ee

We can also integrate by parts in $S_1$.  The boundary terms vanish,
and we can use the equation of motion \eqref{eq:MultiFieldDEOM} to
find
\be\label{eq:S1}
S_1 = -\frac{A_{D-1}}{2} \int_0^\infty dr r^{D-1}
\frac{\partial U}{\partial \phi_i}\phi_i~.
\ee

Our goal is to give, as much as possible, the action of the correct
instanton profile, rather than the action of the actual profile that
we found (which is divergent, because our profile has small
discontinuities where the shooting regions join).  So we are free to
make any transformations to the action in terms of the exact profile,
and then use the approximate profile in the computation.

Computation using $\partial_r\phi_i$ would be a poor choice, because
the derivative is less accurately represented than the function itself
in the numerical result of solving the differential equation.  We
might use \eqref{eq:Derrick} to write the action in terms of $S_2$
alone, but this fails when $D=2$.  So instead we use \eqref{eq:Derrick}
to write the action in terms of $S_1$ alone, and calculate that with
\eqref{eq:S1}.  As long as the profile can be differentiated
analytically, using $\partial U/\partial\phi_i$ does not introduce any
additional error.

Thus we perform numerically the integral
\bel{action3}
	S[\bphi]= -\frac{\pi ^{D/2}}{\Gamma (1+D/2)} \int_0^\infty dr r^{D-1} 
\left[\frac{\partial U}{\partial \phi_i}(\bphi(r))-\frac{\partial
    U}{\partial \phi_i}(\bphifv)\right]\phi_i(r)~.
\ee
With exact computation, $\partial U/\partial \phi_i$ would vanish at
the false vacuum.  But numerical error can give a tiny result, which we subtract off
explicitly. This prevents it from being amplified into an infinite error 
under the integral out to $r = \infty$.

\section{Examples of potentials solved by the code}
\label{sec:Examples}
\begin{figure}
   \centering
   \includegraphics[width=2.7in]{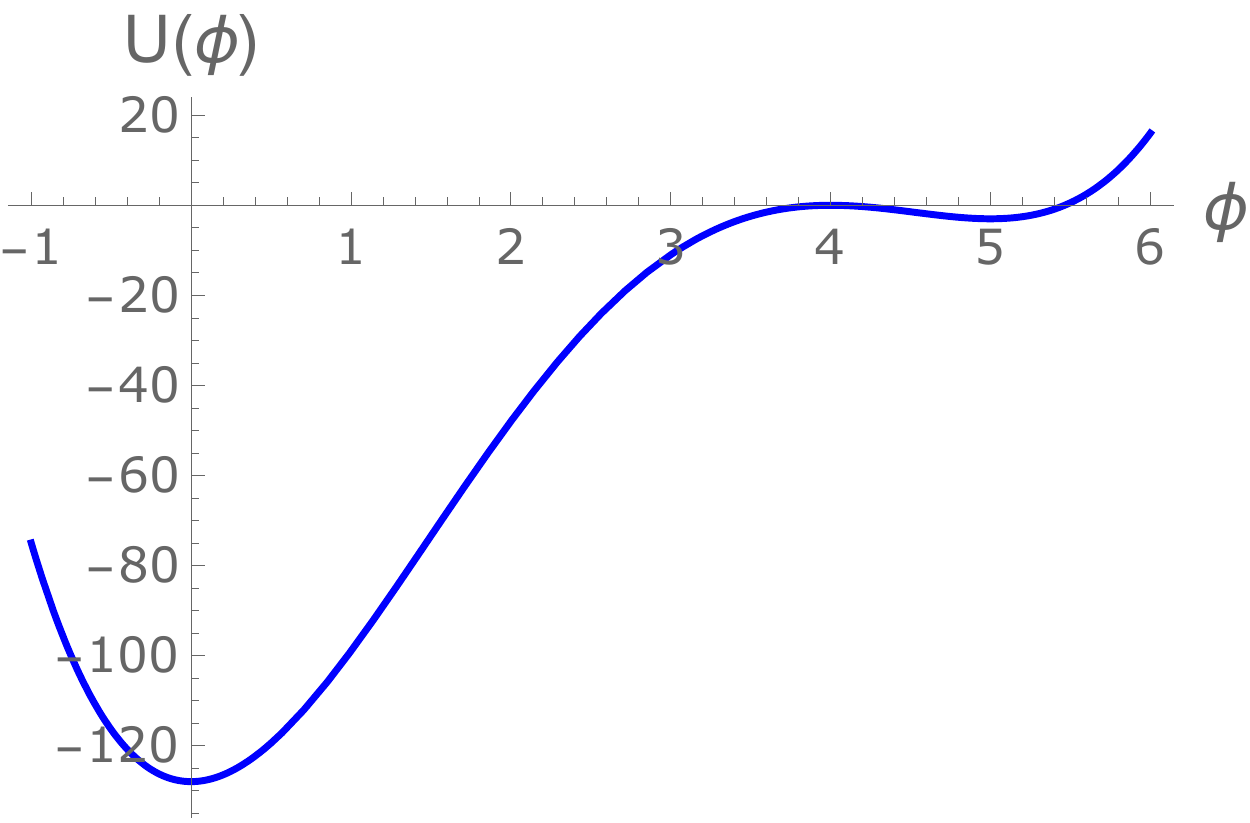} 
   \includegraphics[width=2.7in]{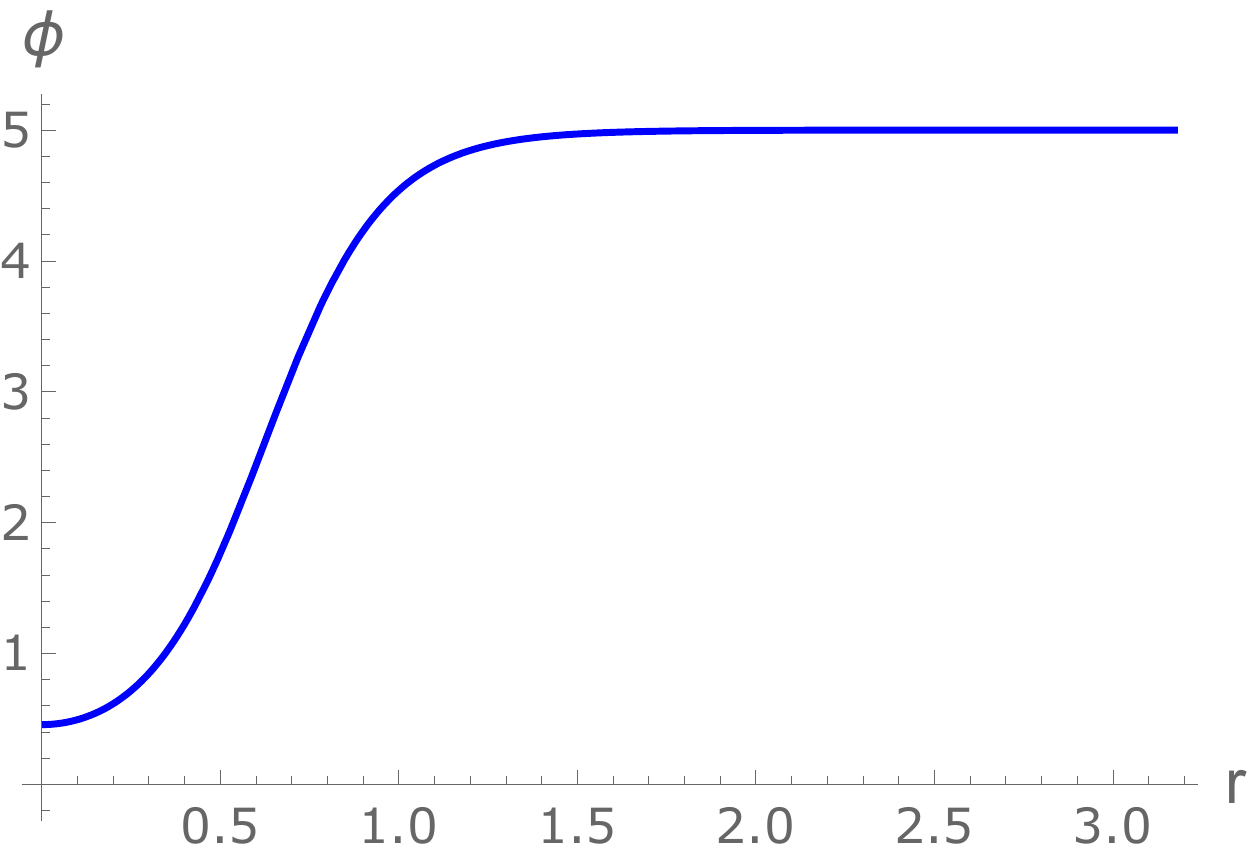} 
   \caption{The potential and field profile for  the one-dimensional potential given in \eqref{1dExample}. This is an example of  a thick-wall solution solved in under four seconds. }
   \label{fig:1dExample}
\end{figure}

Here we show examples of one, two and four field cases solved using
this code.  We tried the one-dimensional potential
\bel{1dExample}
	U(\phi)=  \phi^4 - 12 \phi^3 +40 \phi^2~.
\ee
This potential has minima at $\phi = 0$ and $\phi = 5$, and the code calculated the bounce in under 4 seconds on a Macbook Pro. 
The potential and solution are shown in Figure~\ref{fig:1dExample}.

\begin{figure}
   \centering
   \includegraphics[width=2.7in]{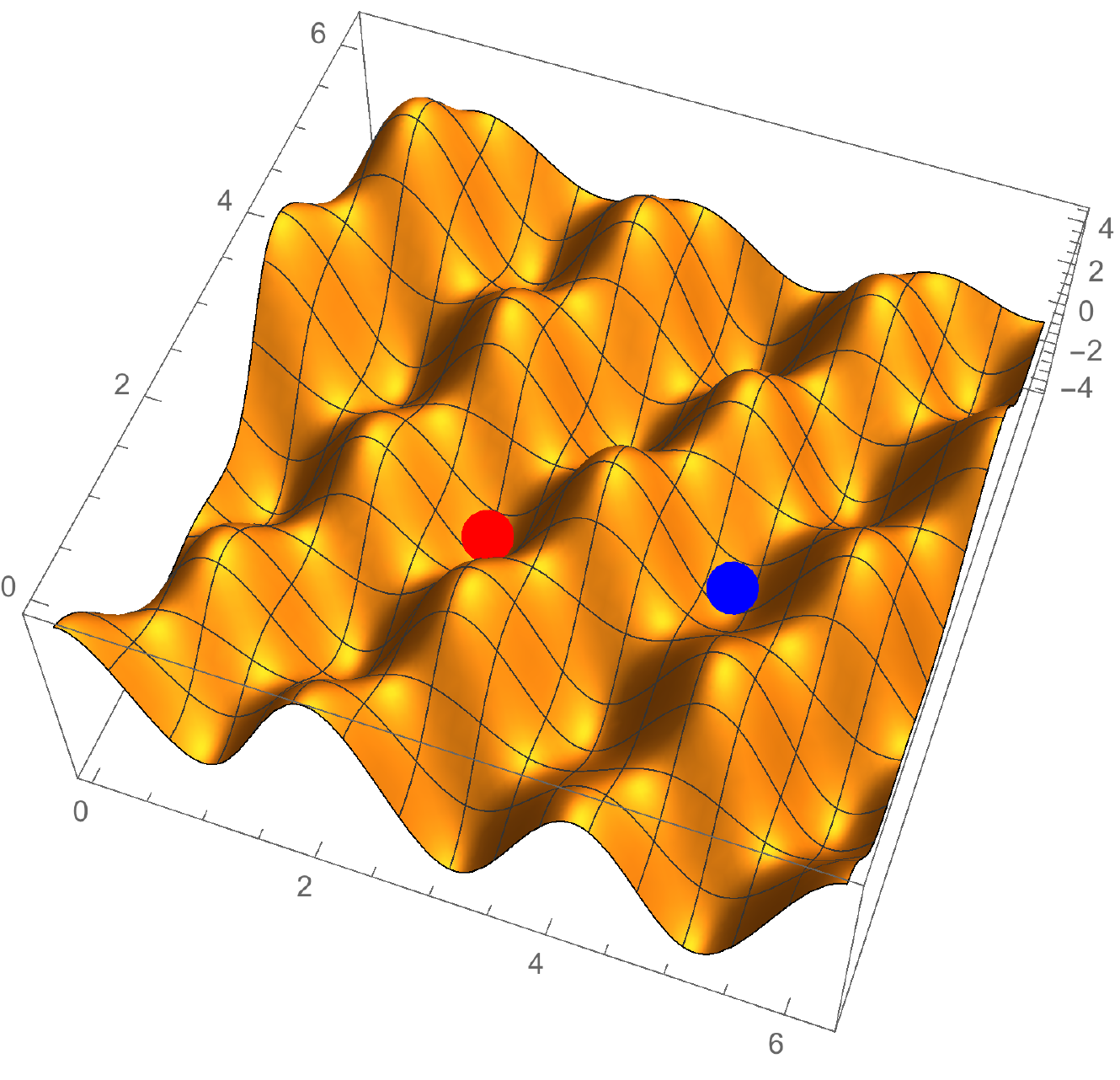} 
   \includegraphics[width=2.7in]{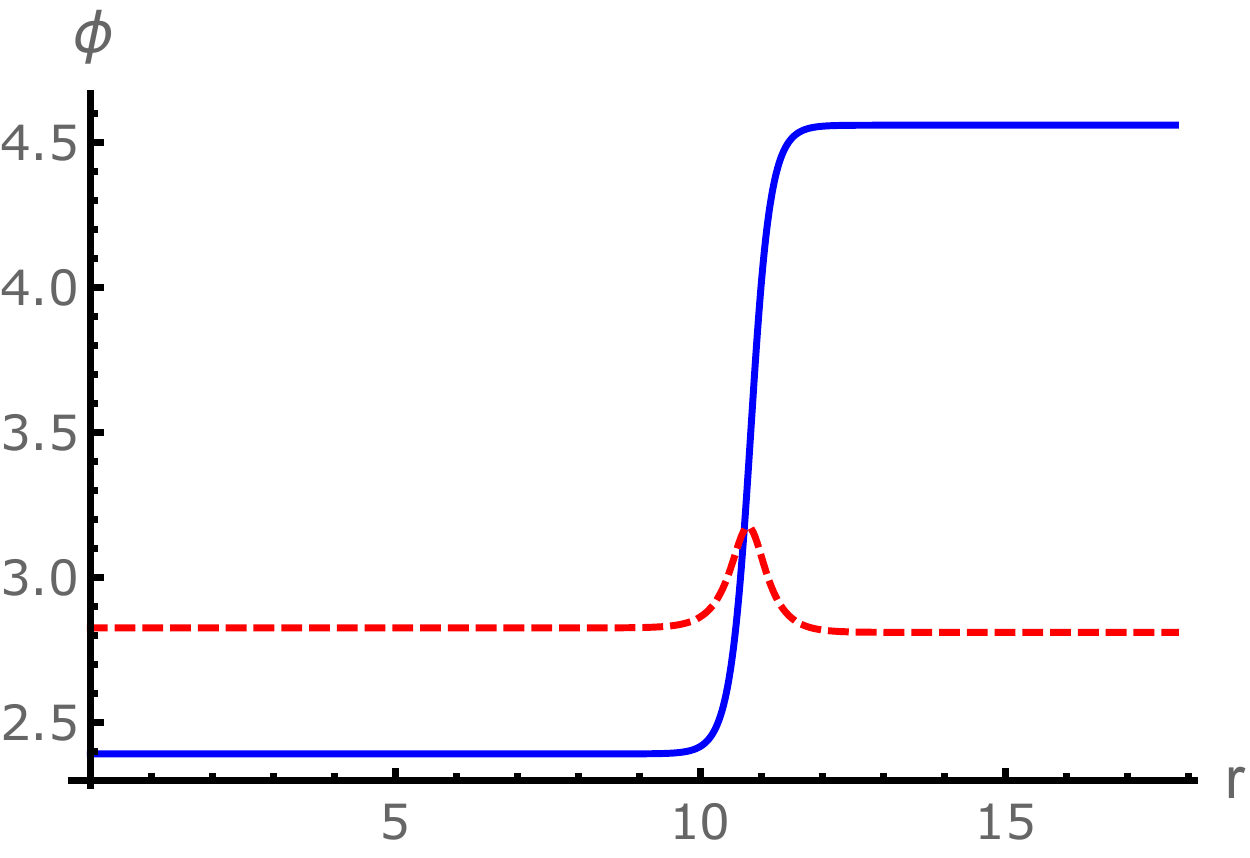} 
   \caption{The potential and field profile for  the two-dimensional potential given in \eqref{twoDExample}.  True and false vacua are shown by the red and blue circles. Blue solid and red dashed lines correspond to $\phi_1$ and $\phi_2$. The true and false vacua are at (2.39338, 2.82768) and (4.56086, 2.81235). }
   \label{fig:2dExample}
\end{figure}

As a two-dimensional example, we solved the bounce solution for the two-field potential given by 
\bel{twoDExample}
	U(\phi_1,\phi_2)=\sin \left(\phi _1-\phi _2\right)+\frac12 \cos \left(\phi _1+\phi _2\right)+\cos 3\left( \phi _1+\phi _2\right)+2 \cos\frac32 \left(2\phi _1- \phi _2\right)~.
\ee
This was solved in ten seconds on a Macbook Pro.  The potential and field profiles are shown in Figure~\ref{fig:2dExample}.

\begin{figure}
   \centering
   \includegraphics[width=2.7in]{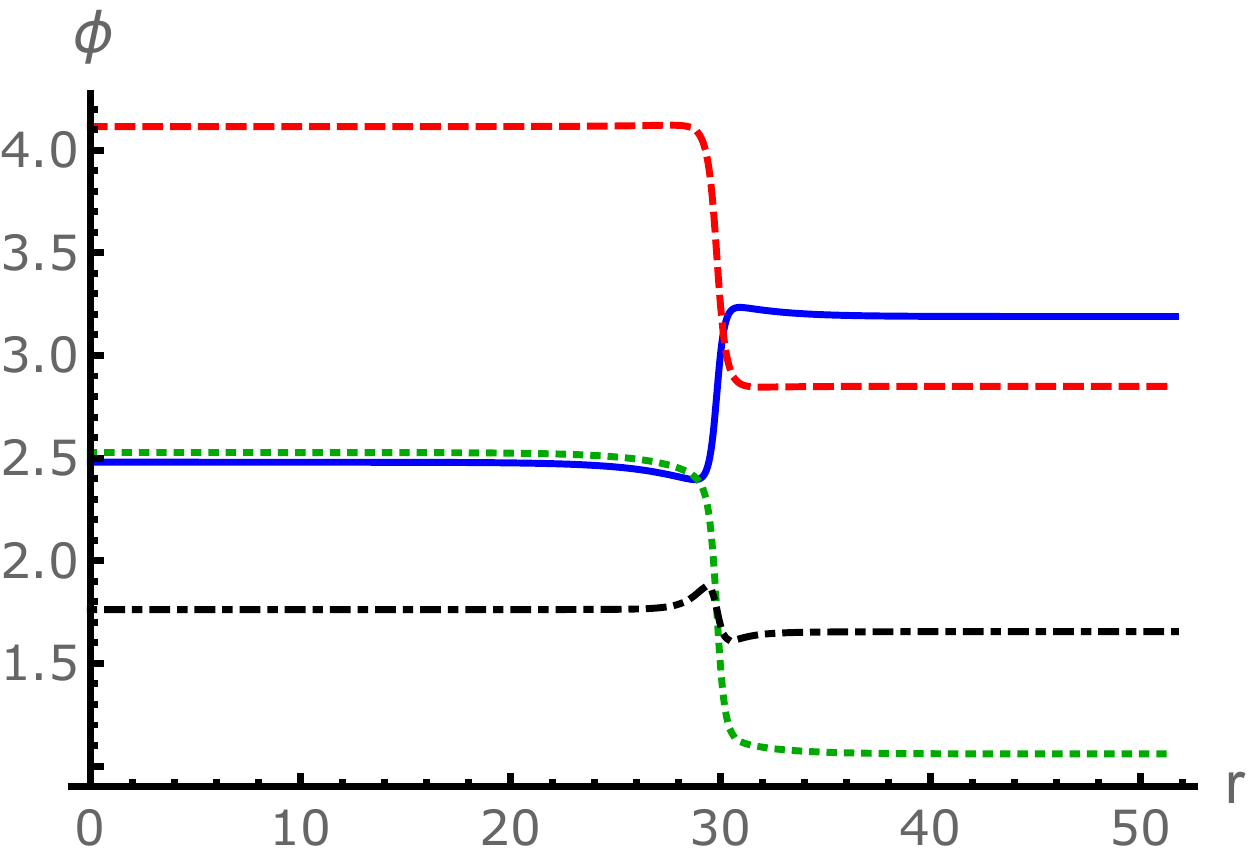} 
   \caption{The potential and field profile for  the four-dimensional potential given in \eqref{4DExample}.  This is an example of a thin-wall solution. The blue solid, red dashed, green dotted and black dot-dashed  correspond to $\phi_1, \phi_2, \phi_3$ and $\phi_4$.}
   \label{fig:4dExample}
\end{figure}

As the last example we show a four-field case which took about fifteen minutes to get the solution. The potential is given by 
\beal{4DExample}
	U&=&-1.2 \cos (\phi _1+2 \phi _2-\phi _3-\phi _4)-1.25 \cos (2
        \phi _1-\phi _2-2 \phi _3-\phi _4)\nn
&&-0.75 \cos (\phi _1-2 \phi _2-2 \phi _3-\phi _4)\nn
	&&-\cos (\phi _1+\phi _2-\phi _3+\phi _4)-0.5 \cos(\phi _1-\phi _2-\phi _3-2 \phi _4)~.
\eea
We found the instanton that gives the tunneling from a false vacuum at
(3.18979, 2.84979, 1.05933, 1.65474) to  a true vacuum at (2.48005, 4.1149, 2.52733, 1.76197). The field profiles are shown in Figure~\ref{fig:4dExample}.

\section{Quality of the code}
\label{sec:Quality}

We describe here more systematically the abilities of the code.

\subsection{Range of application}
We have successfully tested our code on random quartic potentials for
$N$ up to $9$.  We have found bubbles whose radius is larger than the
wall thickness by a factor of $200$, and extreme thick wall cases,
where there is no clear distinction between the interior and the wall.

For comparison, Kusenko \cite{Kusenko:1996jn} computed actions for 8
fields, while Dasgupta \cite{Dasgupta:1996qu} used 10 but computed
only a bound on the action.  As far as we know no other authors have
used more than 3 fields.

\subsection{Success fraction}
We tested our code with potentials in a box of side length $L$.  We
summed up to 10 Fourier modes randomly chosen among those with with
wavelengths down to $L/6$, with random coefficients chosen from
a Gaussian distribution.  For each potential we found all minima and
calculated the action to tunnel from each minimum to the closest
minimum, if that minimum had a smaller energy.

For $N=2$, the code succeeded 96\% of the time in about 10,000 tries,
with a typical time about 10 seconds on a Xeon X5675 processor.  For
$N=3$, the code succeeded in 95\% of about 3,000 tries, with a
typical time of about 1 minute.

\subsection{Speed}
Our code finds solutions in a matter of seconds for one to three
fields, minutes for four to seven fields, and hours for eight to nine
fields.  We have not attempted to run our code for ten or more
fields.

Most of the time is spent computing the Jacobian. The Jacobian for
each shooting region has $4N^2$ elements.  These are determined by
integrating a matrix differential equation, as discussed in appendix
\ref{sec:Jacobian}, which requires matrix multiplication, and thus
$8N^3$ operations for each integration step in the differential
equation.  One could perhaps abandon explicit calculation of the
Jacobian in favor of finite differences, or write a special-purpose
program for doing the integration.  The code could also be run in
parallel, at a minimum by handling the different shooting regions on
different processors.

\subsection{Quality of solution}
Our code supports computations with arbitrary precision numbers, so
numerical error can be made as small as desired at the price of
significantly increased runtime.  Unfortunately, there is still an
inaccuracy resulting from the initial and final analytic regions,
which turns out to be the most significant one.  It may be made small
by decreasing the parameters $f_s$ and $f_e$ (defined in
section~\ref{sec:Readjustingvalues}), but the resulting
parameterization of the problem using $\varbphi_1$ close to
$\varbphi_0$ and $\varbphi_n$ close to $\bphifv$ leads to slow
convergence toward the solution.  Nevertheless, by comparing solutions
with one choice of $f_s$ and $f_e$ to more accurate solutions with a
smaller choice, we can find out how good a result we can get.  The
default choice of $10^{-2}$ for these parameters gives relative
accuracy about $10^{-3}$ in the action, while reducing them to
$10^{-3}$ gives relative accuracy about $10^{-5}$.

For comparison, Ref.~\cite{Kusenko:1996jn, Cline:1999wi} claimed
accuracies of order $10^{-2}$, and \cite{Wainwright:2011kj} claimed
profiles accurate to $10^{-3}$.  Konstandin \& Huber
\cite{Konstandin:2006nd} report errors of $7\times 10^{-6}$ using
$1600$ lattice sites.

\subsection{Possible issues}
\label{sec:PossibleIssues}

Here we describe certain difficulties the user may encounter with the
code.  Some of these may be resolved in future releases.

\begin{itemize}

\item Runaway potentials (i.e., those whose true vacua do not exist)
  are not currently supported.  Note that it is always possible to
  modify such potentials, moving the true vacuum in to finite field
  values, in a way which does not affect the bounce instanton.

\item As discussed in subsection~\ref{subsec:Analytic}, near the
  beginning and end of the profile, we use analytic solutions to the
  potential expanded to second-order in a Taylor series around
  $\varbphi_1$ and $\bphifv$, respectively.  For this to work,
  Mathematica has to be able to compute the second derivative, which
  causes trouble with certain non-analytic functions such as ${\tt
    Abs}$ and ${\tt UnitStep}$.  At the moment the best strategy is to
  smooth out such jumps by analytic functions such as ${\tt Tanh}$.
  The potential should not have jumps of this kind between
  $\varbphi_0$ and $\varbphi_1$ or between $\varbphi_n$ and $\bphifv$,
  or the result will be quite inaccurate.

\item Potentials whose Hessian has a vanishing eigenvalue at the false
  vacuum are not supported.  A vanishing eigenvalue at a point chosen
  as $r_1$ would also be problematic, but this is unlikely ever to
  occur unless the Hessian eigenvalue vanishes over a
  non-infinitesimal region.

\item As discussed above, while the accuracy of the solution can be
  improved by using small values for $f_i$ and $f_s$, the resulting
  parameterization leads to much slower convergence.

\end{itemize}

\section{Comparison with other methods}
\label{sec:Comparison}

To our knowledge, all techniques for finding instanton solutions
involve either shooting or the division of the space of $r$ to be
considered into a lattice, i.e., a sequence of fixed $r$ points.
Some methods use a hybrid of these two techniques.

In simple shooting, one just varies $\varbphi_0$ to look for a
solution which stays near $\bphifv$ for as long as possible.  As
discussed above, this technique is unstable, but it works fine for
thick-wall solutions, where there is not much exponential growth.
Even in thin-wall cases, it is possible to beat the instability by the
use of high precision arithmetic, but the needed precision can be very
high, so this is not an efficient technique.  Avoiding such problems
by using multiple shooting regions gives the algorithm described in
this paper.

Functional minimization problems can be solved by relaxation, in which
one creates a discrete approximation to the function and finds the
minimum by varying the parameters of this approximation.
Unfortunately, the present problem is not minimization of the action,
because the action is a maximum with regard to dilation of the
profile, and a minimum over all other variations.  However, the
solution can be found minimizing an improved action
\cite{Kusenko:1996jn} or the integral of the square of the deviation
from the equations of motion \cite{Moreno:1998bq, John:1998ip}.

Alternatively, Refs.~\cite{Dasgupta:1996qu, Cline:1999wi,
  Wainwright:2011kj} separated out the problem of finding the path
taken by the field in field space, which is strictly a minimization
problem, and solved the one-dimensional problem of finding the profile
by shooting techniques.  Refs.~\cite{Cline:1998rc, Cline:1999wi}
discuss a gradient descent/ascent method for finding stationary points
which are not minima, but according to Ref.~\cite{Cline:1999wi},
splitting the path and profile problems is much more efficient.
Konstandin \& Huber \cite{Konstandin:2006nd} used a lattice equation
of motion, which they solved as a Newton's method problem with two
equations (field and derivative) for each lattice site.

The difficulty of all such methods is that accurate results require
using a large number of points.  Ref.~\cite{Konstandin:2006nd}
reported relative error of $7\times 10^{-6}$ in the action, but they
used 1600 lattice sites to do it.  Such methods are especially
problematic for a large number of fields $N$, because the set of
variables to be operated on is then multiplied by $N$.  The
multi-shooting method described here is able to achieve good accuracy
without a large number of variables.\footnote{Of course this method
  works by solving differential equations, which means that functions
  are evaluated at some sequence of points.  However, the runtime
  grows only linearly with the number of points used in differential
  equation solving, whereas other equation-solving or minimization
  procedures normally have much faster growth with the number of
  variables.}

\section{Conclusion}
\label{sec:Conclusion}
We have presented here a multi-shooting method for the calculation of
instantons for the nucleation of bubbles of true vacuum inside a false
vacuum in a potential depending on several fields.  This method
succeeds nearly all the time in the tests that we have done.  For one
to three fields it runs quite quickly, and even for nine fields it
succeeds, albeit slowly.

We have made our Mathematica implementation of this method freely
available, as described in appendix \ref{sec:UsingTheCode} below.  We
hope it will be useful, both for the calculation of tunneling rates in
specific models, and for statistical surveys of large numbers of
random potentials.

\acknowledgments
We thank Brian Greene, Alex Vilenkin, Erick Weinberg, and I-Sheng Yang
for helpful discussions and comments.  This work was supported
in part by the National Science Foundation under grants 1213888 and
1518742.

\appendix

\section{Calculation of Jacobian}
\label{sec:Jacobian}

In order to implement Powell's hybrid method we need to know how the
equations change when we change the parameters.  One way to do so is
by changing the parameters by small amounts to see how the result
changes.  In fact, Powell \cite{Powell1} gives a procedure to do this,
primarily using the sequence of parameter choices.  However, direct
calculation of the Jacobian is more reliable and less prone to
numerical errors, and it is possible to do that here.

\subsection{Solving for Jacobian along the path}
\label{sec:Jacobian1}
To compute the Jacobian, we need to know, for each shooting step, the
way in which the solution to the differential equation, and thus the
value of the field and its derivative at the endpoint, depend on the
initial conditions.

We start with a system of first order differential equations,
\be\label{eq:1stOrderDiff}	
\fb'(r) = \bF(r,\fb(r))~,
\ee
with the initial conditions $\fb(0)$ given.  That is, $\fb(r)$ is a
vector function of $r$ whose derivative depends upon $r$ and the value
of $\fb$ at r.  We want to know how the solution of the differential equation at some
point $r$ depends on changes to the initial conditions.  So suppose
that we start with infinitesimally different initial conditions
$\bg(0)$.  Assuming $\bF$ is a smooth function, these new
conditions will give rise to an infinitesimally different solution
$\bg(r)$ and we can define $\bA(r) = \bg(r) - \fb(r)$.  Then
$\bA$ satisfies a simple differential equation,
\be\label{eq:diffEqn2}
	\bA'(r) = \bg'(r) - \fb'(r) = \bF(r,\bg) - \bF(r,\fb)
	 = \bF(r,\fb+\bA) - \bF(r,\fb)
        = \frac{\partial \bF}{\partial f_j} A_j(r)~.
\ee

The first-order effect on $f_i(r)$ due to changes in $f_i(0)$ is given
by the Jacobian matrix with components
 \be\label{eq:JacobianDef}
	J_{ij}=\frac{\partial f_i(r)}{\partial f_j(0)}~.
 \ee
To find such a component, we choose $A_j(0) = 1$ for the $j$ of
interest and 0 for the others, and then $J_{ij}(r) = A_i(r)$.  Thus,
for each $j$, the $J_{ij}(r)$ are the solution to the system of
differential equations
\be\label{eq:J1}
	J_{ij}'(r) = \frac{\partial F_i}{\partial f_k} J_{kj}(r)
\quad \text{with initial conditions $J_{lj}(0) = \delta_{lj}$}~.
\ee
Since \eqref{eq:J1} holds for each $j$, we can consider it to be a
matrix differential equation giving all components of $J_{ij}(r)$.

Let us apply this technique to the Jacobian for the differential
equations given in \eqref{eq:MultiFieldDEOM}. First we convert these
equations into a set of first order equations by defining auxiliary
fields $\phi_i'= \phi_{i+N}$ and rewriting \eqref{eq:MultiFieldDEOM}
as
 \begin{eqnarray} \label{eq:1stOrderDiff3}
	\phi_{i+N}'&=& -\frac{D-1}{r} \phi_{i+N} + \frac{\partial U}{\partial \phi_j}~, \cr 
	\phi_i'&=& \phi_{i+N}~.
 \end{eqnarray}
Now the problem is mapped into the form of \eqref{eq:1stOrderDiff} and we get a simple expression 
\be\label{eq:EvolutionMatrix}
          \frac{\partial F_A}{\partial f_B}= \left( 
	  \begin{array}{c|c}
		\mathbf{0} & \mathbf{I} \\ 
                \hline
		{\partial^2 U/\partial \phi_i \partial \phi_j}& -\frac{D-1}{r} \mathbf{I}
	\end{array} 
	  \right)
\ee
 To calculate the Jacobian we first solve the differential equations
 \eqref{eq:MultiFieldDEOM} and then evaluate
 \eqref{eq:EvolutionMatrix} along the solutions $\bphi$ to solve
 \eqref{eq:J1} and get $J_{AB}(r)$.

For the first and last interval we must also take account of the fact that
we specify only $\phi_1$ and $\phi_n$, and $\phi'_1$ and $\phi'_n$ are
determined from these.  Those derivatives then in turn determine the
initial conditions used in shooting.  Thus the Jacobian for these
cases must include the effect of a change in the field value upon the
derivative found by the procedures of section~\ref{subsec:Analytic}.

When we consider variations of $\phi_1$, we should consider changes to
the Hessian matrix $B_{ij}$ of equation \eqref{eq:ABDef}, because of
being evaluated at a different $\bphi$.  But these depend on the third
derivatives of the potential and are complicated to calculate.  So we
have ignored this effect in computing the Jacobian.  This should not
creating any larger problem than ignoring these same third derivatives
in computing the analytic approximations.

\subsection{Jacobian with respect to $\Delta$}

As described in subsection~\ref{sec:AddingNewVar} we do not evaluate
the fields at constant values of $r$. Instead we have an extra
variable $\Delta$ which changes the value of the $r_i$ where we
apply the parameters. Fortunately, we do not need to solve extra
differential equations to compute the effect of changing
$\Delta$.  The matching points are
 \be \label{Ri}
 R_i=\{r_1+\Delta, r_2+\Delta, \ldots r_n+\Delta\}
 \ee
Here we show the procedure for case of $n=4$ case, which is depicted
in figure~\ref{fig:schematics}, with a single field.  Generalization
to arbitrary $n$ and $N$ is straightforward.  First we calculate the
following Jacobian using the technique described in section
\ref{sec:Jacobian1}:
 \be \label{Jacobian1}
 	J=\left(\begin{array}{ccccc}
		\frac{\partial \phi_L(r_2)}{\partial \phi(r_1)}  & 0 & 0 & 0 \\ 
		\frac{\partial \phi'_L(r_2)}{\partial \phi(r_1)}  & 0 & 0 & 0 \\ 
		0 & \frac{\partial \phi_L(r_3)}{\partial \phi(r_2)}  & \frac{\partial \phi_L(r_3)}{\partial \phi'(r_2)}  & - \frac{\partial \phi_R(r_3)}{\partial \phi(r_4)} \\ 
		0 & \frac{\partial \phi'_L(r_3)}{\partial \phi(r_2)}  & \frac{\partial \phi'_L(r_3)}{\partial \phi'(r_2)}  & - \frac{\partial \phi'_R(r_3)}{\partial \phi(r_4)} \\ 
	\end{array}
	\right)~.
 \ee

When we add the new variable $\Delta$, our Jacobian will the same
number of rows, but one more column, giving the dependence of the
various parameters on $\Delta$,
\be \label{Jacobian2}
 	J_{\text{total}}=\left(\begin{array}{cccc|c}
		&&& & \frac{\partial \phi_L(r_2)}{\partial \Delta} \\ 
		&\qquad J \qquad&
                && \frac{\partial \phi'_L(r_2)}{\partial \Delta}\\ 
		&&& & \frac{\partial \phi_L(r_3)}{\partial \Delta} - \frac{\partial \phi_R(r_3)}{\partial \Delta} \\ 
		&&& & \frac{\partial \phi'_L(r_3)}{\partial \Delta} - \frac{\partial \phi'_R(r_3)}{\partial \Delta} \\ 
	\end{array}
	\right)~.
 \ee
There are two types of effects due to $\Delta$.  First, after we
solve the differential equation, changing $r_i$ changes the place at
which the solution is evaluated.  Second, changing $r_i$ changes which
solution of the differential equation we use when we start from the
same initial data $\phi(r_i)$ and $\phi'(r_i)$.  We can propagate
the new data back to the former location of $r_i$, changing the
$\phi$ and $\phi'$ values used as initial conditions appropriately.

In all, we find six types of terms that we need to calculate:
\bea\label{eq:Jtotal}
	\frac{\partial \phi_L(r_3)}{\partial \Delta}&=&   \phi'_L(r_3)-\phi'(r_2)  \frac{\partial \phi_L(r_3)}{\partial \phi(r_2)}  - \phi''(r_2)  \frac{\partial \phi_L(r_3)}{\partial \phi'(r_2)}~, \nn
	\frac{\partial \phi'_L(r_3)}{\partial \Delta}&=&  \phi''_L(r_3)-\phi'(r_2)  \frac{\partial \phi'_L(r_3)}{\partial \phi(r_2)}  - \phi''(r_2)  \frac{\partial \phi'_L(r_3)}{\partial \phi'(r_2)} ~, \nn
	\frac{\partial \phi_L(r_2)}{\partial \Delta}&=&  \phi'_L(r_2) -\phi'(r_1)  \frac{\partial \phi_L(r_2)}{\partial \phi(r_1)}~, \nn
	\frac{\partial \phi'_L(r_2)}{\partial \Delta}&=&  \phi''_L(r_2) -\phi'(r_1)  \frac{\partial \phi'_L(r_2)}{\partial \phi(r_1)}~,\nn
	\frac{\partial \phi_R(r_3)}{\partial \Delta}&=&  \phi'_R(r_3) -\phi'(r_4)  \frac{\partial \phi'_R(r_3)}{\partial \phi(r_4)}~,\nn
	\frac{\partial \phi'_R(r_3)}{\partial \Delta}&=&  \phi''_R(r_3) -\phi'(r_4)  \frac{\partial \phi'_R(r_3)}{\partial \phi(r_4)}~.
\eea 
The partial derivatives on the right in \eqref{eq:Jtotal} are all
elements of the original Jacobian $J$.  The first derivatives at $r_2$
and $r_3$ are variables, while those at $r_1$ and $r_4$ can be
computed from the analytic forms given in
section~\ref{subsec:Analytic}.  Finally, the second derivatives
can be computed using the equations of motion,
\be
	\phi''(r_i)= U'\left(\phi(r_i)\right)-\frac{D-1}{r_i} \phi'(r_i).
\ee

\section{Powell's method for solving equations}
\label{sec:Powell}

M. J. D. Powell \cite{Powell1} describes a method for solving multiple
simultaneous equations that is a hybrid of Newton's method and gradient
descent.  We describe this method briefly here; for more details, see
Ref.~\cite{Powell1}.

Powell's hybrid method attempts to iteratively find the simultaneous
root of a set of nonlinear functions $f_i(x_1, \ldots)$.  Given
some current guess $\bx$, the method finds some step $\bd$ such that
$\bx+\bd$ is a better approximation to the desired root.  This is done
by (potentially) combining two possible steps.

We first linearize the equations around the point $\bx$.  Newton's
method works by considering next the solution to these linearized
equations.  This method is quadratically convergent if one is close
enough to the actual solution.  But if not, the next guess can be much
further from the actual solution than the previous one.

An alternative is to move in the direction which most rapidly decreases
the squared error $F(\bx) = \sum _i x_i^2$.  This is a quadratic in the
linearized $f_i(\bx)$, so it achieves a minimum at some distance from
$\bx$ and we will call this direction and distance the gradient step.

Powell's hybrid method combines these ideas as follows.  It maintains at
each time a desired step size $\Delta$ (in the Euclidean norm on the
space of parameter values).  If the step recommended by Newton's method
is shorter than $\Delta$, Powell's method tries to take the Newton step.
If not, consider the gradient step.  If it is longer than $\Delta$, try
going distance $\Delta$ in the gradient direction.  If it is shorter,
try a linear combination of the gradient step and the Newton step that
has length $\Delta$.

In all cases, if the trial step decreases $F$ we take it, and adjust
the step size based on how much progress we are making.  If the trial
fails to decrease $F$, we try shorter and shorter steps until we find
one that does.  If only an infinitesimal step decreases $F$, it appears
that we have found a local minimum and must give up.

The method starts always with a gradient step.  If the method appears to
be working, it moves more and more toward Newton's method, and
eventually converges rapidly on the solution.  If things are going
poorly, it slowly descends the surface of $F$ until it finds an area
where it can move rapidly towards the solution (or fails at a local
minimum).

In order to apply this method, we need the Jacobian $\partial
f_i/\partial x_j$.  Powell \cite{Powell1} describes techniques
for maintaining an approximate Jacobian by a series of function
evaluations.  But here we are able to compute the Jacobian numerically,
as described in appendix~\ref{sec:Jacobian}, so we do not need these
techniques.
As described in section~\ref{sec:AddingNewVar}, we extended the method to
allow more variables than functions.  This has no effect at all on the
gradient method: we move in the direction that most rapidly decreases
$F$, just as before.  For Newton's method, there are now many possible
steps that solve the linearized equations, and we choose the shortest of
these.

\section{Obtaining and running the code}
\label{sec:UsingTheCode}
We have implemented the above procedures in Mathematica.  The code can
be downloaded from \url{http://cosmos.phy.tufts.edu/AnyBubble/}.
There is a manual which explains how to find instanton solutions
and all the options available in the code.

\bibliography{anybubble}

\providecommand{\href}[2]{#2}\begingroup\raggedright\begin{thebibliography}{10}

\bibitem{Langer:1969bc}
J.~S. Langer, \emph{{Statistical theory of the decay of metastable states}},
  \href{http://dx.doi.org/10.1016/0003-4916(69)90153-5}{\emph{Annals Phys.}
  {\bf 54} (1969) 258--275}.

\bibitem{Leggett:1975te}
A.~J. Leggett, \emph{{A theoretical description of the new phases of liquid
  He-3}}, \href{http://dx.doi.org/10.1103/RevModPhys.47.331}{\emph{Rev. Mod.
  Phys.} {\bf 47} (1975) 331--414}.

\bibitem{Cabibbo:1979ay}
N.~Cabibbo, L.~Maiani, G.~Parisi and R.~Petronzio, \emph{{Bounds on the
  Fermions and Higgs Boson Masses in Grand Unified Theories}},
  \href{http://dx.doi.org/10.1016/0550-3213(79)90167-6}{\emph{Nucl. Phys.} {\bf
  B158} (1979) 295--305}.

\bibitem{Hung:1979dn}
P.~Q. Hung, \emph{{Vacuum Instability and New Constraints on Fermion Masses}},
  \href{http://dx.doi.org/10.1103/PhysRevLett.42.873}{\emph{Phys. Rev. Lett.}
  {\bf 42} (1979) 873}.

\bibitem{EliasMiro:2011aa}
J.~Elias-Miro, J.~R. Espinosa, G.~F. Giudice, G.~Isidori, A.~Riotto and
  A.~Strumia, \emph{{Higgs mass implications on the stability of the
  electroweak vacuum}},
  \href{http://dx.doi.org/10.1016/j.physletb.2012.02.013}{\emph{Phys. Lett.}
  {\bf B709} (2012) 222--228}, [\href{http://arxiv.org/abs/1112.3022}{{\tt
  1112.3022}}].

\bibitem{Dasgupta:1996qu}
I.~Dasgupta, \emph{{Estimating vacuum tunneling rates}},
  \href{http://dx.doi.org/10.1016/S0370-2693(96)01685-1}{\emph{Phys. Lett.}
  {\bf B394} (1997) 116--122}, [\href{http://arxiv.org/abs/hep-ph/9610403}{{\tt
  hep-ph/9610403}}].

\bibitem{Sarid:1998sn}
U.~Sarid, \emph{{Tools for tunneling}},
  \href{http://dx.doi.org/10.1103/PhysRevD.58.085017}{\emph{Phys. Rev.} {\bf
  D58} (1998) 085017}, [\href{http://arxiv.org/abs/hep-ph/9804308}{{\tt
  hep-ph/9804308}}].

\bibitem{Steinhardt:1981ct}
P.~J. Steinhardt, \emph{{Relativistic Detonation Waves and Bubble Growth in
  False Vacuum Decay}},
  \href{http://dx.doi.org/10.1103/PhysRevD.25.2074}{\emph{Phys. Rev.} {\bf D25}
  (1982) 2074}.

\bibitem{Witten:1984rs}
E.~Witten, \emph{{Cosmic Separation of Phases}},
  \href{http://dx.doi.org/10.1103/PhysRevD.30.272}{\emph{Phys. Rev.} {\bf D30}
  (1984) 272--285}.

\bibitem{Hogan:1986qda}
C.~J. Hogan, \emph{{Gravitational radiation from cosmological phase
  transitions}}, {\emph{Mon. Not. Roy. Astron. Soc.} {\bf 218} (1986)
  629--636}.

\bibitem{Kosowsky:1992rz}
A.~Kosowsky, M.~S. Turner and R.~Watkins, \emph{{Gravitational waves from first
  order cosmological phase transitions}},
  \href{http://dx.doi.org/10.1103/PhysRevLett.69.2026}{\emph{Phys. Rev. Lett.}
  {\bf 69} (1992) 2026--2029}.

\bibitem{Cline:1999wi}
J.~M. Cline, G.~D. Moore and G.~Servant, \emph{{Was the electroweak phase
  transition preceded by a color broken phase?}},
  \href{http://dx.doi.org/10.1103/PhysRevD.60.105035}{\emph{Phys. Rev.} {\bf
  D60} (1999) 105035}, [\href{http://arxiv.org/abs/hep-ph/9902220}{{\tt
  hep-ph/9902220}}].

\bibitem{Randall:2006py}
L.~Randall and G.~Servant, \emph{{Gravitational waves from warped spacetime}},
  \href{http://dx.doi.org/10.1088/1126-6708/2007/05/054}{\emph{JHEP} {\bf 05}
  (2007) 054}, [\href{http://arxiv.org/abs/hep-ph/0607158}{{\tt
  hep-ph/0607158}}].

\bibitem{Kuzmin:1985mm}
V.~A. Kuzmin, V.~A. Rubakov and M.~E. Shaposhnikov, \emph{{On the Anomalous
  Electroweak Baryon Number Nonconservation in the Early Universe}},
  \href{http://dx.doi.org/10.1016/0370-2693(85)91028-7}{\emph{Phys. Lett.} {\bf
  B155} (1985) 36}.

\bibitem{Witten:1981gj}
E.~Witten, \emph{{Instability of the Kaluza-Klein Vacuum}},
  \href{http://dx.doi.org/10.1016/0550-3213(82)90007-4}{\emph{Nucl. Phys.} {\bf
  B195} (1982) 481--492}.

\bibitem{Gross:1982cv}
D.~J. Gross, M.~J. Perry and L.~G. Yaffe, \emph{{Instability of Flat Space at
  Finite Temperature}},
  \href{http://dx.doi.org/10.1103/PhysRevD.25.330}{\emph{Phys. Rev.} {\bf D25}
  (1982) 330--355}.

\bibitem{Guth:2007ng}
A.~H. Guth, \emph{{Eternal inflation and its implications}},
  \href{http://dx.doi.org/10.1088/1751-8113/40/25/S25}{\emph{J. Phys.} {\bf
  A40} (2007) 6811--6826}, [\href{http://arxiv.org/abs/hep-th/0702178}{{\tt
  hep-th/0702178}}].

\bibitem{Vilenkin:2006xv}
A.~Vilenkin, \emph{{A Measure of the multiverse}},
  \href{http://dx.doi.org/10.1088/1751-8113/40/25/S22}{\emph{J. Phys.} {\bf
  A40} (2007) 6777}, [\href{http://arxiv.org/abs/hep-th/0609193}{{\tt
  hep-th/0609193}}].

\bibitem{Bousso:2000xa}
R.~Bousso and J.~Polchinski, \emph{{Quantization of four form fluxes and
  dynamical neutralization of the cosmological constant}},
  \href{http://dx.doi.org/10.1088/1126-6708/2000/06/006}{\emph{JHEP} {\bf 06}
  (2000) 006}, [\href{http://arxiv.org/abs/hep-th/0004134}{{\tt
  hep-th/0004134}}].

\bibitem{Susskind:2003kw}
L.~Susskind, \emph{{The Anthropic landscape of string theory}},
  \href{http://arxiv.org/abs/hep-th/0302219}{{\tt hep-th/0302219}}.

\bibitem{Aazami:2005jf}
A.~Aazami and R.~Easther, \emph{{Cosmology from random multifield potentials}},
  \href{http://dx.doi.org/10.1088/1475-7516/2006/03/013}{\emph{JCAP} {\bf 0603}
  (2006) 013}, [\href{http://arxiv.org/abs/hep-th/0512050}{{\tt
  hep-th/0512050}}].

\bibitem{Sarangi:2007jb}
S.~Sarangi, G.~Shiu and B.~Shlaer, \emph{{Rapid Tunneling and Percolation in
  the Landscape}},
  \href{http://dx.doi.org/10.1142/S0217751X09042529}{\emph{Int. J. Mod. Phys.}
  {\bf A24} (2009) 741--788}, [\href{http://arxiv.org/abs/0708.4375}{{\tt
  0708.4375}}].

\bibitem{Tye:2007ja}
S.~H.~H. Tye, \emph{{A Renormalization Group Approach to the Cosmological
  Constant Problem}},  \href{http://arxiv.org/abs/0708.4374}{{\tt 0708.4374}}.

\bibitem{Huang:2008jr}
Q.-G. Huang and S.~H.~H. Tye, \emph{{The Cosmological Constant Problem and
  Inflation in the String Landscape}},
  \href{http://dx.doi.org/10.1142/S0217751X0904316X}{\emph{Int. J. Mod. Phys.}
  {\bf A24} (2009) 1925--1962}, [\href{http://arxiv.org/abs/0803.0663}{{\tt
  0803.0663}}].

\bibitem{Greene:2013ida}
B.~Greene, D.~Kagan, A.~Masoumi, D.~Mehta, E.~J. Weinberg et~al.,
  \emph{{Tumbling through a landscape: Evidence of instabilities in
  high-dimensional moduli spaces}},
  \href{http://dx.doi.org/10.1103/PhysRevD.88.026005}{\emph{Phys.Rev.} {\bf
  D88} (2013) 026005}, [\href{http://arxiv.org/abs/1303.4428}{{\tt
  1303.4428}}].

\bibitem{Aravind:2014pva}
A.~Aravind, B.~S. DiNunno, D.~Lorshbough and S.~Paban, \emph{{Analyzing
  multifield tunneling with exact bounce solutions}},
  \href{http://dx.doi.org/10.1103/PhysRevD.91.025026}{\emph{Phys. Rev.} {\bf
  D91} (2015) 025026}, [\href{http://arxiv.org/abs/1412.3160}{{\tt
  1412.3160}}].

\bibitem{Dine:2015ioa}
M.~Dine and S.~Paban, \emph{{Tunneling in Theories with Many Fields}},
  \href{http://dx.doi.org/10.1007/JHEP10(2015)088}{\emph{JHEP} {\bf 10} (2015)
  088}, [\href{http://arxiv.org/abs/1506.06428}{{\tt 1506.06428}}].

\bibitem{Masoumi:2016eqo}
A.~Masoumi and A.~Vilenkin, \emph{{Vacuum statistics and stability in axionic
  landscapes}},
  \href{http://dx.doi.org/10.1088/1475-7516/2016/03/054}{\emph{JCAP} {\bf 1603}
  (2016) 054}, [\href{http://arxiv.org/abs/1601.01662}{{\tt 1601.01662}}].

\bibitem{Kobzarev:1974cp}
I.~{\relax Yu}. Kobzarev, L.~B. Okun and M.~B. Voloshin, \emph{{Bubbles in
  Metastable Vacuum}}, {\emph{Sov. J. Nucl. Phys.} {\bf 20} (1975) 644--646}.

\bibitem{Coleman:1977py}
S.~R. Coleman, \emph{{The Fate of the False Vacuum. 1. Semiclassical Theory}},
  \href{http://dx.doi.org/10.1103/PhysRevD.15.2929,
  10.1103/PhysRevD.16.1248}{\emph{Phys.Rev.} {\bf D15} (1977) 2929--2936}.

\bibitem{Linde:1981zj}
A.~D. Linde, \emph{{Decay of the False Vacuum at Finite Temperature}},
  \href{http://dx.doi.org/10.1016/0550-3213(83)90293-6,
  10.1016/0550-3213(83)90072-X}{\emph{Nucl. Phys.} {\bf B216} (1983) 421}.

\bibitem{Bayliss}
A.~Bayliss, \emph{A double shooting scheme for certain unstable and singular
  boundary value problems}, {\emph{Mathematics of Computation} {\bf 32} (1978)
  61}.

\bibitem{Wainwright:2011kj}
C.~L. Wainwright, \emph{{CosmoTransitions: Computing Cosmological Phase
  Transition Temperatures and Bubble Profiles with Multiple Fields}},
  \href{http://dx.doi.org/10.1016/j.cpc.2012.04.004}{\emph{Comput. Phys.
  Commun.} {\bf 183} (2012) 2006--2013},
  [\href{http://arxiv.org/abs/1109.4189}{{\tt 1109.4189}}].

\bibitem{Morrison:1962:MSM:355580.369128}
D.~D. Morrison, J.~D. Riley and J.~F. Zancanaro, \emph{Multiple shooting method
  for two-point boundary value problems},
  \href{http://dx.doi.org/10.1145/355580.369128}{\emph{Commun. ACM} {\bf 5}
  (Dec., 1962) 613--614}.

\bibitem{Konstandin:2006nd}
T.~Konstandin and S.~J. Huber, \emph{{Numerical approach to multi dimensional
  phase transitions}},
  \href{http://dx.doi.org/10.1088/1475-7516/2006/06/021}{\emph{JCAP} {\bf 0606}
  (2006) 021}, [\href{http://arxiv.org/abs/hep-ph/0603081}{{\tt
  hep-ph/0603081}}].

\bibitem{Olum:2016svf}
K.~D. Olum and A.~Masoumi, \emph{{Auxiliary variables for nonlinear equations
  with softly broken symmetries}},  \href{http://arxiv.org/abs/1611.03950}{{\tt
  1611.03950}}.

\bibitem{Derrick:1964ww}
G.~H. Derrick, \emph{{Comments on nonlinear wave equations as models for
  elementary particles}}, \href{http://dx.doi.org/10.1063/1.1704233}{\emph{J.
  Math. Phys.} {\bf 5} (1964) 1252--1254}.

\bibitem{Kusenko:1996jn}
A.~Kusenko, P.~Langacker and G.~Segre, \emph{{Phase transitions and vacuum
  tunneling into charge and color breaking minima in the MSSM}},
  \href{http://dx.doi.org/10.1103/PhysRevD.54.5824}{\emph{Phys. Rev.} {\bf D54}
  (1996) 5824--5834}, [\href{http://arxiv.org/abs/hep-ph/9602414}{{\tt
  hep-ph/9602414}}].

\bibitem{Moreno:1998bq}
J.~M. Moreno, M.~Quiros and M.~Seco, \emph{{Bubbles in the supersymmetric
  standard model}},
  \href{http://dx.doi.org/10.1016/S0550-3213(98)00283-1}{\emph{Nucl. Phys.}
  {\bf B526} (1998) 489--500}, [\href{http://arxiv.org/abs/hep-ph/9801272}{{\tt
  hep-ph/9801272}}].

\bibitem{John:1998ip}
P.~John, \emph{{Bubble wall profiles with more than one scalar field: A
  Numerical approach}},
  \href{http://dx.doi.org/10.1016/S0370-2693(99)00272-5}{\emph{Phys. Lett.}
  {\bf B452} (1999) 221--226}, [\href{http://arxiv.org/abs/hep-ph/9810499}{{\tt
  hep-ph/9810499}}].

\bibitem{Cline:1998rc}
J.~M. Cline, J.~R. Espinosa, G.~D. Moore and A.~Riotto, \emph{{String mediated
  electroweak baryogenesis: A Critical analysis}},
  \href{http://dx.doi.org/10.1103/PhysRevD.59.065014}{\emph{Phys. Rev.} {\bf
  D59} (1999) 065014}, [\href{http://arxiv.org/abs/hep-ph/9810261}{{\tt
  hep-ph/9810261}}].

\bibitem{Powell1}
M.~J.~D. Powell, \emph{A hybrid method for nonlinear equations},  in
  \emph{Numerical methods for nonlinear algebraic equations} (P.~Rabinowitz,
  ed.), ch.~6, pp.~87--114.
\newblock Gordon and Breach Science Publishers, New York, 1970.

\end{thebibliography}\endgroup

\end{document}